\documentclass[12pt,draftcls,onecolumn]{IEEEtran}
\usepackage[applemac]{inputenc}
\usepackage{epsfig}
\usepackage{latexsym}
\usepackage{amsfonts}
\usepackage{amssymb}
\usepackage{amsmath}
\usepackage{amsbsy}
\usepackage{color,graphicx}
\usepackage{multirow}
\usepackage{epic}
\usepackage{algorithm}
\usepackage{algpseudocode,MnSymbol,ifsym}
\usepackage{amssymb,amsmath}
\usepackage{url}
\usepackage{dsfont}

\newtheorem{proposition}{\textbf{Proposition}}

\newtheorem{lemma}{Lemma}
\newtheorem{remark}{Remark}

\newcommand{\dv}{\mathbf} 
\newcommand{\mc}{\mathcal} 
\newcommand{\qed}{\hfill \ensuremath{\Box}}

\graphicspath{{./}{./fig/}}

\begin{document}

\fontencoding{OT1}\fontsize{9.4}{11.25pt}\selectfont
\title{Compute-and-Forward on a Multiaccess Relay Channel: Coding and Symmetric-Rate Optimization}
\vspace{1cm}

\author{\vspace{0cm}
\authorblockN{ \small Mohieddine El Soussi \qquad Abdellatif Zaidi \qquad
Luc Vandendorpe\thanks{The material in this paper has been presented in part at the IEEE International Conference on Communication, Ottawa, Canada, June 2012. This work has been supported in part by the IAP BESTCOM project funded by BELSPO, and by the SCOOP project. }
\thanks{Mohieddine El Soussi and Luc Vandendorpe are with ICTEAM, Universit\'e catholique de Louvain, Place du Levant, 2, 1348 Louvain-la-Neuve, Belgium. Email: \{mohieddine.elsoussi,luc.vandendorpe\}@uclouvain.be}
\thanks{Abdellatif Zaidi is with Universit\'e Paris-Est Marne La Vall\'ee,
77454 Marne la Vall\'ee Cedex 2, France. Email:
abdellatif.zaidi@univ-mlv.fr}}}

\vspace{1cm}

\maketitle

\begin{abstract}

We consider a system in which two users communicate with a destination with the help of a half-duplex relay. Based on the compute-and-forward scheme, we develop and evaluate the performance of coding strategies that are of network coding spirit. In this framework, instead of decoding the users' information messages, the destination decodes two integer-valued linear combinations that relate the transmitted codewords. Two decoding schemes are considered. In the first one, the relay computes one of the linear combinations and then forwards it to the destination. The destination computes the other linear combination based on the direct transmissions. In the second one, accounting for the side information available at the destination through the direct links, the relay compresses what it gets using Wyner-Ziv compression and conveys it to the destination. The destination then computes the two linear combinations, locally. For both coding schemes, we discuss the design criteria, and derive the allowed symmetric-rate. Next, we address the power allocation and the selection of the integer-valued coefficients to maximize the offered symmetric-rate; an iterative coordinate descent method is proposed. The analysis shows that the first scheme can outperform standard relaying techniques in certain regimes, and the second scheme, while relying on feasible structured lattice codes, can at best achieve the same performance as regular compress-and-forward for the multiaccess relay network model that we study. The results are illustrated through some numerical examples.
\end{abstract}


\section{Introduction}\label{secI}

\IEEEPARstart{N}{etwork} coding was introduced by Ahlswede \textit{et al.} in \cite{ACLY00} for wired networks. It refers to each intermediate node sending out a function of the packets that it receives, an operation which is more general than simple routing \cite{FF56,EFS56}. In linear network coding, intermediate nodes compute and send out linear combinations over an appropriate finite field of the packets that they receive. In general, the function does not need to be linear. Although they are generally suboptimal for general wireline networks, linear network codes have been shown optimum for multicasting \cite{LYC03,KM03}. Moreover they have appreciable features, in particular simplicity (e.g., see \cite{YLCZ05,FS07} are references therein). For these reasons, most of the research on network coding has focused on linear codes. 

The development of efficient network coding techniques for wireless networks is more involved than for wired network coding, essentially because of fading, interference and noise effects. For general wireless networks, the quantize-map-and-forward scheme of \cite{ADT11} and the more general noisy network coding scheme of \cite{H-LKGC11} can be seen as interesting and efficient extensions for wireless settings of the original network coding principle. However, quantize-map-and-forward and noisy network coding are based on random coding arguments. For wireless networks, efficient linear network coding techniques make use of structured codes, and in particular lattices \cite{CS88}. Lattices play an important role in network coding for diverse network topologies, such as the two-way relay channel \cite{NWS07,KMT08}, the Gaussian network \cite{NG11a} and others.

Recently, Nazer and Gastpar propose and analyse a scheme in which receivers decode finite-field linear combinations of transmitters' messages, instead of the messages themselves. The scheme is called "Compute-and-forward" (CoF) \cite{NG11a}, and can be implemented with or without the presence of relay nodes. In this setup, a receiver that is given a sufficient number of linear combinations recovers the transmitted messages by solving a system of independent linear equations that relate the transmitted symbols. Critical in this scheme, however, is that the coefficients of the equations to decode must be integer-valued. This is necessitated by the fact that a combination of codewords should itself be a codeword so that it be decodable. Lattice codes have exactly this property, and are thus good candidates for implementing compute-and-forward.

Compute-and-forward is a promising scheme for network coding in wireless networks. However, the problem of selecting the integer coefficients optimally, i.e., in a manner that allows to recover the sent codewords from the decoded equations and, at the same time, maximizes the transmission rate is not an easy task. As shown by Nazer and Gastpar \cite{NG11a}, the compute-and-forward optimally requires a match between the channel gains and the desired integer coefficients. However, in real communication scenarios, it is unlikely that the channels would produce gains that correspond to integer values. This problem has been addressed in \cite{NSGS09}, where the authors develop a superposition strategy to mitigate the non-integer channel coefficients penalty. The selection of which integer combinations to decode is then a crucial task to be performed by the receivers. While it can be argued that linear combinations that are recovered at the same physical entity can always be chosen appropriately, i.e., in a way enabling system inversion to solve for the sent codewords, selecting these linear combinations in a distributed manner, i.e., at physically separate nodes, is less easy to achieve. By opposition to previous works, part of this paper focuses on this issue.

In this work, we consider communication over a two-user multiaccess relay channel. In this model, two independent users communicate with a destination with the help of a common relay node, as shown in Figure~\ref{SystemModel}. The relay is assumed to operate in half-duplex mode.

\subsection{Contributions}\label{secI_subsecA}

We establish two coding schemes for the multiaccess relay model that we study. The first coding scheme is based on compute-and-forward at the relay node. The relay uses what it receives from the transmitters during the first transmission period to compute a linear combination with integer coefficients of the users' codewords. It then sends this combination to the destination during the second transmission period. In addition to the linear combination that it gets from the relay's transmission, the destination recovers the required second linear combination from what it gets directly from the transmitters, through the direct links. If the set of integer coefficients that are selected at the relay and the destination are chosen appropriately, the destination can solve for the transmitted codewords. We consider individual power constraints at the transmitters and the relay, and analyse the symmetric-rate offered by this coding scheme \cite{SZV11} \cite{SZV12}. 

In the second coding scheme both required linear integer combinations of the users' codewords are recovered locally at the destination. More specifically, the relay quantizes its output from the users' transmission during the first transmission period using Wyner-Ziv compression \cite{WZ76}. In doing so, it accounts for the output at the destination during this transmission period as available side information at the decoder. Then, the relay sends the lossy version of its output to the destination during the second transmission period. The destination determines the two required linear combinations, as follows. It utilizes an appropriate combination of the output from the users' transmission during the first period and of the compressed version of the relay's output during the second period; from this combination, two independent linear combinations relating the users' codewords are recovered.

For the two coding schemes, we target the optimization of the transmitters and the relay powers, and of the integer coefficients of the linear combinations to maximize the achievable symmetric-rate. These optimization problems are NP hard. For the two coding schemes, we develop an iterative approach that finds the appropriate power and integer coefficients alternately. More specifically, we show that the problem of finding appropriate integer coefficients for a given set of powers has the same solution as an approximated mixed integer quadratic programming (MIQP) problem with quadratic constraints. Also, we show that the problem of finding the appropriate power policy at the transmitters and the relay for a given set of integer coefficients is a non-linear non-convex optimization problem. We formulate and solve this problem through geometric programming and a successive convex approximation approach \cite{CTPOJ07}. 

Our analysis shows that, for certain regimes, i.e., channel conditions, the first scheme outperforms known strategies for this model that do not involve forms of network coding, such as those based on having the relay implements classic amplify-and-forward (AF), decode-and-forward (DF) or compress-and-forward (CF) relaying schemes. The second scheme offers rates that are at best as large as those offered by compress-and-forward for the multiaccess relay network that we study. However, this scheme relies on feasible structured lattice codes and utilizes linear receivers, and so, from a practical viewpoint it offers advantages over standard CF which is based on random binning arguments. We illustrate our results by means of some numerical examples. The analysis also shows the benefit obtained from allocating the powers and the integer coefficients appropriately.

\subsection{Outline and Notation}\label{secI_subsecB}

An outline of the remainder of this paper is as follows. Section~\ref{secII} describes in more details the communication model that we consider in this work. It also contains some preliminaries on lattices and known results from the literature for the setup under consideration where the relay uses standard techniques. In Section~\ref{secIII}, we describe our coding strategies and analyse the symmetric rates that are achievable using these strategies. Section~\ref{secIV} is devoted to the optimization of the power values and the integer-valued coefficients for an objective function which is the symmetric-rate. Section~\ref{secV} contains some numerical examples, and Section~\ref{secVI} concludes the paper.

We use the following notations throughout the paper. Lowercase boldface letters are used to denote vectors, e.g., $\dv x$. Upper case boldface letters are used to denote matrices, e.g., $\dv X$. Calligraphic letters designate alphabets, i.e., $\mc X$. The cardinality of a set $\mc X$ is denoted by $|\mc X|$. For matrices, we use the notation $\dv X \in \mathbb{R}^{m{\times}n}$, $m, n \in \mathbb{N}$, to mean that $\dv X$ is an $m$-by-$n$ matrix, i.e., with $m$ rows and $n$ columns, and its elements are real-valued. Also, we use $\dv X^T$ to designate the $n$-by-$m$ matrix transpose of $\dv X$; and, if $m=n$, $\text{det}(\dv X)$ to designate the determinant of $\dv X$. We use $\dv I_n$ to denote the $n$-by-$n$ identity matrix; and $\dv 0$ to denote a matrix whose elements are all zeros (its size will be evident from the context). Similarly, for vectors, we write $\dv x \in \mathbb{A}^n$, e.g., $\mathbb{A}=\mathbb{R}$ or $\mathbb{A}=\mathbb{Z}$, to mean that $\dv x$ is a column
  vector of size $n$, and its elements are in $\mathbb{A}$. For a vector $\dv x \in \mathbb{R}^n$, $\|\dv x\|$ designates the norm of $\dv x$ in terms of Euclidean distance; and for a scalar $x \in \mathbb{R}$, $|x|$ stands for the absolute value of $x$, i.e., $|x|=x$ if $x \geq 0$ and $|x|=-x$ if $x \leq 0$. For two vectors $\dv x \in \mathbb{R}^n$ and $\dv y \in \mathbb{R}^n$, the vector $\dv z = \dv x \circ \dv y \in \mathbb{R}^n$ denotes the Hadamard product of $\dv x$ and $\dv y$, i.e., the vector whose $i$th element is the product of the $i$th elements of $\dv x$ and $\dv y$, i.e., $z_i=(\dv x \circ \dv y)_i=x_iy_i$. Also, we use $\text{Var}(\dv x)$ to denote the power of $\dv x$ i.e. $(1/n)\mathbb{E}[||\dv x||^2]$. The Gaussian distribution with mean $\mu$ and variance $\sigma^2$ is denoted by $\mc N(\mu, \sigma^2)$. Finally, throughout the paper except where otherwise mentioned, logarithms are taken to base $2$; and, for $x \in \mathbb{R}$, $\log^{+}(x) := \max\{\log(x)
 , 0\}$.

\section{Preliminaries and System Model}\label{secII}

In this section, we first recall some basics on lattices, and then present the system model that we study and recall some known results from the literature, obtained through classic relaying, i.e., amplify-and-forward, decode-and-forward and compress-and-forward. The results given in Section~\ref{secII_subsecC} will be used later for comparison purposes in this paper.

\subsection{Preliminaries on Lattices}\label{secII_subsecA}

Algebraically, an $n$-dimensional lattice $\Lambda$ is a discrete additive subgroup of $\mathbb{R}^n$. Thus, if $\boldsymbol{\lambda}_1 \in \Lambda$ and $\boldsymbol{\lambda}_2 \in \Lambda$, then $(\boldsymbol{\lambda}_1 +\boldsymbol{\lambda}_2) \in \Lambda$ and $(\boldsymbol{\lambda}_1 - \boldsymbol{\lambda}_2) \in \Lambda$. For an $n$-dimensional lattice $\Lambda \in \mathbb{R}^n$, there exists (at least one) matrix $\dv G \in \mathbb{R}^{n\times n}$ such that any lattice point $\boldsymbol{\lambda} \in \Lambda$ can be written as an integer combination of the columns of $\dv G$. The matrix $\dv G$ is called generator matrix of $\Lambda$, and satisfies 
\begin{equation}
\Lambda = \{\boldsymbol{\lambda} = \mathrm{\mathbf{z}}\mathrm{\mathbf{G}:\mathrm{\mathbf{z}} \in \mathbb{Z}^{n}}\}.
\end{equation}
A lattice quantizer $Q_{\Lambda}$ : $\mathbb{R}^{n} \to \Lambda$ maps a point $\mathrm{\mathbf{x}} \in \mathbb{R}^n$ to the nearest lattice point in Euclidean distance, i.e.,
\begin{equation}
Q_{\Lambda}(\mathrm{\mathbf{x}})=\arg \min_{\boldsymbol{\lambda} \: \in \:\Lambda}\|\mathrm{\mathbf{x}}-\boldsymbol{\lambda}\|.
\end{equation}
The Voronoi region $\mathcal{V}(\boldsymbol{\lambda})$ of $\boldsymbol{\lambda} \in \Lambda$ is the set of all points in $\mathbb{R}^{n}$ that are closer to $\boldsymbol{\lambda}$ than to any other lattice point, i.e.,  
\begin{equation}
\mathcal{V}(\boldsymbol{\lambda})= \{\mathrm{\mathbf{x}} \in \mathbb{R}^n \::\: Q_{\Lambda}(\mathrm{\mathbf{x}})=\boldsymbol{\lambda}\}.
\end{equation}
The \textit{fundamental Voronoi region} $\mc V$ of lattice $\Lambda$ is the Voronoi region $\mathcal{V}(\dv 0)$, i.e., $\mc V=\mathcal{V}(\dv 0)$.
The modulo reduction with respect to $\Lambda$ returns the quantization error, i.e.,
\begin{equation}
[\mathrm{\mathbf{x}}]\:\mathrm{mod}\:\Lambda = \mathrm{\mathbf{x}} -Q_{\Lambda}(\mathrm{\mathbf{x}}) \in \mc V.
\end{equation}
The \textit{second moment} $\sigma^2_{\Lambda}$ quantifies per dimension the average power for a random variable that is uniformly distributed over $\mc V$, i.e.,
\begin{equation}
\sigma^2_{\Lambda} =\frac{1}{n\text{Vol}(\mathcal{V})}\int_{\mathcal{V}}\|\mathrm{\mathbf{x}}\|^2d\mathrm{\mathbf{x}}
\end{equation}
where $\text{Vol}(\mathcal{V})$ is the volume of $\mathcal{V}$. The \textit{normalized second moment} of $\Lambda$ is defined as 
\begin{equation}
G({\Lambda}) =\frac{\sigma^2_{\Lambda}}{\text{Vol}(\mathcal{V})^{2/n}}.
\end{equation}
A lattice $\Lambda$ is said to be nested into another lattice $\Lambda_{\text{FINE}}$ if $\Lambda \subset \Lambda_{\text{FINE}}$, i.e., every point of $\Lambda$ is also a point of $\Lambda_{\text{FINE}}$. We refer to $\Lambda$ as the coarse lattice and to $\Lambda_{\text{FINE}}$ as the fine lattice. Also, given two nested lattices $\Lambda \subset \Lambda_{\text{FINE}}$, the set of all the points of the fine lattice $\Lambda_{\text{FINE}}$ that fall in the fundamental Voronoi region $\mathcal{V}$ of the coarse lattice $\Lambda$ form a codebook 
\begin{eqnarray}
\mathcal{C}^{} = \Lambda_{\text{FINE}} \cap \mathcal{V} = \{\mathrm{\mathbf{x}}= \boldsymbol{\lambda}\:\mathrm{mod}\:\Lambda,\:\boldsymbol{\lambda} \in \Lambda_{\text{FINE}}\}.
\end{eqnarray}
The rate of this codebook is
\begin{equation}
R = \frac{1}{n}\log_2(|\mc C^{}|).
\end{equation}

Finally, the $\mathrm{mod}$ operation satisfies the following properties:
\begin{align}
(\text{P1}) &\:\qquad [[\mathrm{\mathbf{x}}]\:\mathrm{mod}\:\Lambda+\mathrm{\mathbf{y}}]\:\mathrm{mod}\:\Lambda  =  [\mathrm{\mathbf{x}}+\mathrm{\mathbf{y}}]\:\mathrm{mod}\:\Lambda, \qquad \forall \: \dv x \in \mathbb{R}^n, \: \dv y \in \mathbb{R}^n \nonumber\\
(\text{P2}) &\:\qquad  [k([\mathrm{\mathbf{x}}]\:\mathrm{mod}\:\Lambda)]\:\mathrm{mod}\:\Lambda  =  [k\mathrm{\mathbf{x}}]\:\mathrm{mod}\:\Lambda, \qquad \forall \: k \in \mathbb{Z}, \: \dv x \in \mathbb{R}^n\nonumber\\
(\text{P3})&\:\qquad \gamma([\mathrm{\mathbf{x}}]\:\mathrm{mod}\:\Lambda)  =  [\gamma\mathrm{\mathbf{x}}]\:\mathrm{mod}\:\gamma\Lambda, \qquad \forall \: \gamma \in \mathbb{R}, \: \dv x \in \mathbb{R}^n.
\label{properties-modulo-operation}
\end{align}

\subsection{System Model}\label{secII_subsecB}

\begin{figure}[!ht]
  \begin{center}
  \includegraphics[width=0.7\linewidth]{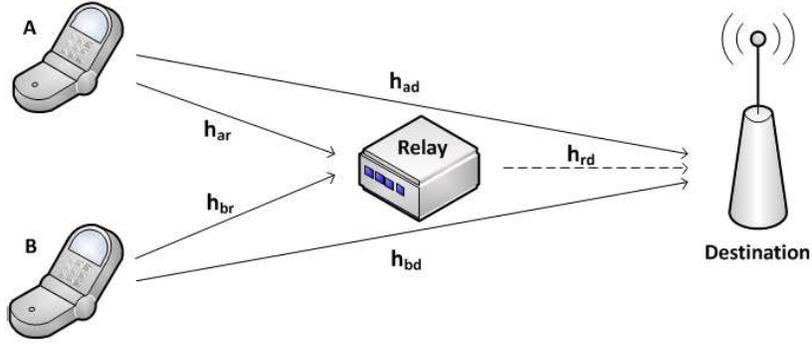}
  \end{center}
  \caption{Multiple-access channel with a half-duplex relay}
  \label{SystemModel}
\end{figure}

We consider the communication system shown in Figure \ref{SystemModel}. Two transmitters $A$ and $B$ communicate with a destination $D$ with the help of a common relay $R$. Transmitter $A$, and Transmitter $B$ want to transmit the messages $W_a \in \mc W_a$, and $W_b \in \mc W_b$ to the destination reliably, in $2n$ uses of the channel. At the end of the transmission, the destination guesses the pair of transmitted messages using its output. Let $R_a$ be the transmission rate of message $W_a$ and $R_b$ be the transmission rate of message $W_b$. We concentrate on the \textit{symmetric} rate case, i.e., $R_a=R_b=R$, or equivalently, $|\mc W_a|=|\mc W_b|=2^{2nR}$. We measure the system performance in terms of the allowed achievable symmetric-rate $R_{\text{sym}} = R_a =R_b = R$. Also, we divide the transmission time into two transmission periods with each of length $n$ channel uses. The relay operates in a half-duplex mode.

During the first transmission period, Transmitter $A$ encodes its message $W_a \in [1,2^{2nR}]$ into a codeword $\dv x_a$ and sends it over the channel. Similarly, Transmitter $B$ encodes its message $W_b \in [1,2^{2nR}]$ into a codeword $\dv x_b$ and sends it over the channel. Let $\dv y_r$ and $\dv y_d$ be the signals received respectively at the relay and at the destination during this period. These signals are given by
\begin{eqnarray}
\mathrm{\mathbf{y}}_{r} &=& h_{ar}\mathrm{\mathbf{x}}_{a} + h_{br}\mathrm{\mathbf{x}}_{b}+ \mathrm{\mathbf{z}}_{r}\nonumber \\
\mathrm{\mathbf{y}}_{d} &=& h_{ad}\mathrm{\mathbf{x}}_{a} + h_{bd}\mathrm{\mathbf{x}}_{b}+ \mathrm{\mathbf{z}}_{d},
\label{outputs-relay-destination-first-transmission-period}
\end{eqnarray}
where \mbox{$h_{ad}$} and \mbox{$h_{bd}$} are the channel gains on the links transmitters-to-destination, \mbox{$h_{ar}$} and \mbox{$h_{br}$} are the channel gains on the links transmitters-to-relay, and $\mathrm{\mathbf{z}}_{r}$ and $\mathrm{\mathbf{z}}_{d}$ are additive background noises at the relay and the destination. 

During the second transmission period, the relay sends a codeword $\mathrm{\mathbf{\tilde{x}}}_{r}$ to help both transmitters. During this period, the destination receives
\begin{eqnarray}
\mathrm{\mathbf{\tilde{y}}}_{d}&=&h_{rd}\mathrm{\mathbf{\tilde{x}}}_{r} + \mathrm{\mathbf{\tilde{z}}}_{d},
\label{output-destination-second-transmission-period}
\end{eqnarray}
where \mbox{$h_{rd}$} is the channel gain on the link relay-to-destination, and $\mathrm{\mathbf{\tilde{z}}}_{d}$ is additive background noise.

Throughout, we assume that all channel gains are real-valued, fixed and known to all the nodes in the network; and the noises at the relay and the destination are independent among each others, and independently and identically distributed (i.i.d) Gaussian, with zero mean and variance $N$. Furthermore, we consider the following individual constraints on the transmitted power (per codeword), 
\begin{align}
\mathbb{E}[\|\dv x_a\|^2]=n\beta^2_aP \leq nP_a, \qquad \mathbb{E}[\|\dv x_b\|^2]=n\beta^2_bP \leq nP_b, \qquad \mathbb{E}[\|\mathrm{\mathbf{\tilde{x}}}_{r}\|^2]=n\beta^2_rP \leq nP_r,
\end{align}
where $P_a \geq 0$, $P_b \geq 0$ and $P_r \geq 0$ are some constraints imposed by the system; $P \geq 0$ is given, and $\beta_a$, $\beta_b$ and $\beta_r$ are some scalars that can be chosen to adjust the actual transmitted powers, and are such that $0 \leq |\beta_a| \leq \sqrt{P_a/P}$, $0 \leq |\beta_b| \leq \sqrt{P_b/P}$ and $0 \leq |\beta_r| \leq \sqrt{P_r/P}$. For convenience, we will sometimes use the shorthand vector notation $\mathrm{\mathbf{h}}_d=[h_{ad},\: h_{bd}]^T$, $\mathrm{\mathbf{h}}_r=[h_{ar},\: h_{br}]^T$ $\in \mathbb{R}^{2}$ and $\boldsymbol{\beta}=[\beta_a,\beta_b, \beta_r]^T$ $\in \mathbb{R}^{3}$, and the shorthand matrix notation $\dv H = [\mathrm{\mathbf{h}}_d^T;\: \mathrm{\mathbf{h}}_r^T]$ $\in \mathbb{R}^{2 \times 2}$ . Also, we will find it useful to sometimes use the notation $\boldsymbol{\beta_s}$ to denote the vector composed of the first two components of vector $\boldsymbol{\beta}$, i.e., $\boldsymbol{\beta_s}=[\beta_a, \beta_b]^T$  -- the subscript ``s" standing for ``sources". Finally, the signal-to-noise ratio will be denoted as $\text{snr} = P/N$ in the linear scale, and by $\text{SNR} = 10\log_{10}(\text{snr})$ in decibels in the logarithmic scale.


\subsection{Symmetric Rates Achievable Through Classic Relaying}\label{secII_subsecC}

In this section, we review some known results from the literature for the model we study. These results will be used for comparisons in Section~\ref{secV}. 
\subsubsection{Amplify-and-Forward}\label{secII_subsecC_subsubsec1}

The relay receives $\dv y_r$ as given by \eqref{outputs-relay-destination-first-transmission-period} during the first transmission period. It simply scales $\dv y_r$ to the appropriate available power and sends it to the destination during the second transmission period. That is, the relay outputs $\mathrm{\mathbf{\tilde{x}}}_{r} = \gamma\:\:\mathrm{\mathbf{y}}_{r}$ , with $\gamma = \sqrt{\beta_r^2\text{snr}/(1+\text{snr}\:\|\boldsymbol{\beta_s} \circ \dv h_r\|^2)}$.
The destination estimates the transmitted messages from its output vectors $(\dv y_d, \mathrm{\mathbf{\tilde{y}}}_{d})$. Using straightforward algebra, it can be shown~\cite{SKM04} that this results in the following achievable sum rate
\begin{eqnarray}
R^{\text{AF}}_{\text{sum}} &=& \max\: \frac{1}{4}\log\Bigg(\text{det}\Big(\mathrm{\mathbf{I}}_{2}+ \beta_a^2\text{snr}(\dv h_a\dv h_a^T) + \beta_b^2\text{snr}(\dv h_b\dv h_b^T)\Big)\Bigg),
\label{achievable-sum-rate-AF}
\end{eqnarray}
where the vectors are given by $\dv h_i=[h_{id},\: h_{ir}h_{rd}\gamma/(\sqrt{1+\gamma^2|h_{rd}|^2})]^T$ for $i=a,b$, and the maximization is over $\boldsymbol{\beta}$.

\noindent The achievable sum rate \eqref{achievable-sum-rate-AF} does not require the two users to transmit at the same rate. Recall that, for a symmetric rate point to be achievable, both transmitters must be able to communicate their messages with at least that rate. Under the constraint of \textit{symmetric-rate}, it can be shown rather straightforwardly \cite{NG11a} that the following symmetric-rate is achievable with the relay operating on the amplify-and-forward mode,
\begin{align}
R^{\text{AF}}_{\text{sym}} = \max\: \frac{1}{4}\min\: \Bigg\{ & \log\left(\text{det}\left(\mathrm{\mathbf{I}}_{2}+ \beta_a^2\text{snr}(\dv h_a\dv h_a^T)\right)\right), \nonumber\\
& \log\left(\text{det}\left(\mathrm{\mathbf{I}}_{2}+ \beta_b^2\text{snr}(\dv h_b\dv h_b^T)\right)\right),\: \frac{1}{2} \log\left(\text{det}\left(\mathrm{\mathbf{I}}_{2}+ \beta_a^2\text{snr}(\dv h_a\dv h_a^T) + \beta_b^2\text{snr}(\dv h_b\dv h_b^T)\right)\right)\Bigg\}.
\label{achievable-sum-rate-AF-symmetric-rate}
\end{align}
%
\subsubsection{Decode-and-Forward}\label{secII_subsecC_subsubsec2}

At the end of the first transmission period, the relay decodes the message pair $(W_a,W_b)$ and then, during the second transmission period, sends a codeword $\mathrm{\mathbf{\tilde{x}}}_{r}$ that is independent of $\dv x_a$ and $\dv x_b$ and carries both messages. The relay employs superposition coding and splits its power among the two messages. It can be shown easily that the resulting achievable sum rate is given by~\cite{SLPM07}
\begin{align}
R^{\text{DF}}_{\text{sum}} = \max\: \frac{1}{4}\min\Bigg\{&\log\left(1+\text{snr}\: \|\boldsymbol{\beta_s} \circ \dv h_r\|^2\right), \log\left(1+\text{snr}\:\|\boldsymbol{\beta_s} \circ \dv h_d\|^2\right)+\log\left(1+\text{snr}\:|h_{rd}|^2\beta_r^2\right)\Bigg\},
\label{achievable-sum-rate-DF}
\end{align}
where the maximization is over $\boldsymbol{\beta}$. Under the constraint of \textit{symmetric-rate}, it can be shown rather straightforwardly \cite{NG11a} that the following symmetric-rate is achievable with the relay operating on the decode-and-forward mode,
\begin{equation}
R^{\text{DF}}_{\text{sym}} =  \max\: \frac{1}{4}\min\: \Bigg\{R(\dv h_r),\:R(\dv h_d)+\frac{1}{2}\log\left(1+ \text{snr}|h_{rd}|^2\beta_r^2\right)\Bigg\},
\label{achievable-sum-rate-DF-symmetric-rate}
\end{equation}
where 
\begin{align}
R(\dv h_i) = \min\:\Bigg\{ & \log\left(1+ \text{snr}|h_{ai}|^2\beta_a^2\right),\: \log\left(1+ \text{snr}|h_{bi}|^2\beta_b^2\right),\: \frac{1}{2} \log\left(1+\text{snr}\: \|\boldsymbol{\beta_s} \circ \dv h_i\|^2\right)\Bigg\}.
\end{align}
\subsubsection{Compress-and-Forward}\label{secII_subsecC_subsubsec3}

At the end of the first transmission period, the relay quantizes the received $\dv y_r$ using Wyner-Ziv compression [14], accounting for the available side information $\dv y_d$ at the destination. It then sends an independent codeword $\mathrm{\mathbf{\tilde{x}}}_{r}$ that carries the compressed version of $\dv y_r$. The destination guesses the transmitted messages using its output from the direct transmission along with the lossy version of the output of the relay that is recovered during the second transmission period. It can be shown that the resulting achievable sum rate is given by~\cite{SLPM07,KGG05},
\begin{eqnarray}
R^{\text{CF}}_{\text{sum}} &=& \max\: \frac{1}{4}R^{\text{CF}},
\label{achievable-sum-rate-CF}
\end{eqnarray}
where
\begin{eqnarray}
R^{\text{CF}} &=& \max\: \log\left(\frac{\left(1+\text{snr}\|\boldsymbol{\beta_s} \circ \dv h_d\|^2\right)\left(1+D/N+\text{snr}\|\boldsymbol{\beta_s} \circ \dv h_r\|^2\right)-\text{snr}^2\big((\boldsymbol{\beta_s}\circ\dv h_r)^T(\boldsymbol{\beta_s}\circ\dv h_d)\big)^2 }{(1+D/N)}\right),
\end{eqnarray}
the maximization is over $\boldsymbol{\beta_s}$ and $D \geq 0$, where $D$ is the distortion due to Wyner-Ziv compression, which is given by,
\begin{eqnarray}
D &=& \frac{N^2\left(1+\text{snr}\|\boldsymbol{\beta_s} \circ \dv h_r\|^2\right)}{|h_{rd}|^2P_r}-\frac{N^2\big(\text{snr}(\boldsymbol{\beta_s}\circ\dv h_r)^T(\boldsymbol{\beta_s}\circ\dv h_d)\big)^2}{|h_{rd}|^2P_r\left(1+\text{snr}\|\boldsymbol{\beta_s} \circ \dv h_d\|^2\right)}.
\label{distortion-achievable-sum-rate-CF}
\end{eqnarray}
Under the constraint of \textit{symmetric-rate}, it can be shown rather easily that the following symmetric-rate is achievable with the relay operating on the compress-and-forward mode,
\begin{align}
R^{\text{CF}}_{\text{sym}} = \max\: \frac{1}{4}\min\: \Bigg\{ &\log\left(1 + \text{snr}|h_{ad}|^2\beta_a^2 + \frac{\text{snr}|h_{ar}|^2\beta_a^2}{1+D/N}\right),\: \log\left(1 + \text{snr}|h_{bd}|^2\beta_b^2 + \frac{\text{snr}|h_{br}|^2\beta_b^2}{1+D/N}\right),\:\frac{1}{2}R^{\text{CF}}\Bigg\}.
\label{achievable-sum-rate-CF-symmetric-rate}
\end{align}

\section{Network Coding Strategies}\label{secIII}

In this section, we develop two coding strategies that are both based on the compute-and-forward strategy of~\cite{NG11a}. The two strategies differ essentially through the operations implemented by the relay. In the first strategy, the relay computes an appropriate linear combination that relates the transmitters' codewords and forwards it to the destination. The destination computes the other required linear combination from what it gets through the direct links. In the second strategy the relay sends a lossy version of its outputs to the destination, obtained through Wyner-Ziv compression \cite{WZ76}. The destination then obtains the desired two linear combinations locally, by using the recovered output from the relay and the output obtained directly from the transmitters. 

\subsection{Compute-and-Forward at the Relay}\label{secIII_subsecA}

The following proposition provides an achievable symmetric-rate for the multiaccess relay model that we study.

\vspace{0.1cm}

\begin{proposition}\label{proposition-achievable-sum-rate-compute-and-forward-at-relay}
For any set of channel vector $\mathrm{\mathbf{h}} = [h_{ar},\: h_{br},\: h_{ad},\: h_{bd},\: h_{rd}]^T \in \mathbb{R}^5$, the following symmetric-rate is achievable for the multiaccess relay model that we study:
\begin{eqnarray}
R^{\text{CoF}}_{\text{sym}} = \max\: \frac{1}{4}\min\left\{\log^{+}\left(\left(\|\mathrm{\mathbf{t}}\|^2-\frac{P((\boldsymbol{\beta_s} \circ \mathrm{\mathbf{h}}_d)^T \mathrm{\mathbf{t}})^2}{N+P\|\boldsymbol{\beta_s} \circ \mathrm{\mathbf{h}}_d\|^2}\right)^{-1}\right)\right.,\:\left.\log^{+}\left(\left(\|\mathrm{\mathbf{k}}\|^2-\frac{P((\boldsymbol{\beta_s} \circ \mathrm{\mathbf{h}}_r)^T \mathrm{\mathbf{k}})^2}{N+P\|\boldsymbol{\beta_s} \circ \mathrm{\mathbf{h}}_r\|^2}\right)^{-1}\right),\: \log\left(1+\frac{P|h_{rd}|^2\beta^2_r}{N}\right)\right\},
\label{achievable-sum-rate-compute-and-forward-at-relay}
\end{eqnarray}
where the maximization is over $\boldsymbol{\beta}$ and over the integer coefficients $\dv k \in \mathbb{Z}^2$ and $\dv t \in \mathbb{Z}^2$ such that $|\text{det}(\dv k, \dv t)| \geq 1$.
\end{proposition}

\vspace{0.1cm}

As we already indicated, in the coding scheme that we use for the proof of Proposition~\ref{proposition-achievable-sum-rate-compute-and-forward-at-relay}, the relay first computes a linear combination with integer coefficients  of the transmitters codewords and then forwards this combination to the destination during the second transmission period. The destination computes another linear combination that relates these codewords using its output from the direct transmissions. With an appropriate choice of the integer-valued coefficients of the combinations, the destination obtain two equations that can be solved for the transmitted codewords.

\textbf{Proof:} In what follows, we first describe the transmission scheme and the encoding procedures at the transmitters and the relay. Then, we describe the decoding procedures at the relay and the destination, and analyze the allowed symmetric-rate. Elements of this proof are similar to that of \cite[Theorem 5]{NG11a}.

Let $\Lambda$ be an $n$-dimensional lattice that is good for quantization in the sense of \cite{EZ04} and whose second moment is equal to $P$, i.e., $\sigma^2_{\Lambda}=P$. We denote by $G(\Lambda)$ and $\mc V$ respectively the normalized second moment and the fundamental Voronoi region of lattice $\Lambda$. Also, let $\Lambda_{\text{FINE}} \supseteq \Lambda$ be a lattice that is good for AWGN in the sense of \cite[Definition 23]{NG11a}, and chosen such that the codebook $\mc C^{} = \Lambda_{\text{FINE}} \cap \mathcal{V}$ be of cardinality $2^{2nR}$ \cite{ZSE02}. We designate by $\mc V_{\text{FINE}}$ the fundamental Voronoi region of lattice $\Lambda_{\text{FINE}}$. The coarse lattice $\Lambda$ and the fine lattice $\Lambda_{\text{FINE}}$ form a pair of nested lattices that we will utilize as a structured code. The rate (per-channel use) of this code is given by the logarithm of the nesting ratio, i.e.,
\begin{equation}
R=\frac{1}{2n}\log\Big(\frac{\text{Vol}(\mc V)}{\text{Vol}(\mc V_{\text{FINE}})}\Big).
\end{equation}
\noindent Let $\dv k=[k_a, k_b] \in \mathbb{Z}^2$ and $\dv t=[t_a, t_b] \in \mathbb{Z}^2$ be given such that $|\text{det}(\dv k, \dv t)| = |k_at_b-k_bt_a| \geq 1$. 

\noindent The encoding and transmission scheme is as follows. 

\textit{Encoding:} Let $(W_a,W_b)$ be the pair of messages to be transmitted. Let $\dv u_a$, $\dv u_b$ and $\dv u_r$ be some dither vectors that are drawn independently and uniformly over $\mc V$ and known by all nodes in the network. Since the codebook $\mc C^{}$ is of size $2^{2nR}=|\mc W_a|$, there exists a one-to-one mapping function $\phi_a(\cdot)$ between the set of messages $\{W_a\}$ and the nested lattice code $\mc C^{}$. Similarly, there exists a one-to-one mapping function $\phi_b(\cdot)$ between the set of messages $\{W_b\}$ and the nested lattice code $\mc C^{}$. Let $\dv v_a=\phi_a(W_a)$ and $\dv v_b=\phi_b(W_b)$, where $\dv v_a \in \mc C^{}$ and $\dv v_b \in \mc C^{}$.

During the first transmission period, to transmit message $W_a$, Transmitter $A$ sends 
\begin{equation}
\mathrm{\mathbf{x}}_a = \beta_a\left(\left[\mathrm{\mathbf{v}}_a - \mathrm{\mathbf{u}}_a \right]\:\mathrm{mod}\:\Lambda \right),
\label{input-first-transmitter-first-strategy}
\end{equation}
 for some $\beta_a \in \mathbb{R}$ such that $0 \leq |\beta_a| \leq \sqrt{P_a/P}$; and to transmit message  $W_b$, Transmitter $B$ sends
\begin{equation}
\mathrm{\mathbf{x}}_b = \beta_b\left(\left[\mathrm{\mathbf{v}}_b - \mathrm{\mathbf{u}}_b \right]\:\mathrm{mod}\:\Lambda \right),
\label{input-second-transmitter-first-strategy}
\end{equation}
where $0 \leq |\beta_b| \leq \sqrt{P_b/P}$. The scalars $\beta_a$ and $\beta_b$ are chosen so as to adjust the transmitters' powers during this period.

\noindent As will be shown shortly the relay decodes correctly an integer combination $\dv e_2=k_a{\dv v_a}+k_b{\dv v_b}$ from what it receives during the first transmission period. It then sends
\begin{equation}
\mathrm{\mathbf{\tilde{x}}}_{r} = \beta_r\left(\left[k_{a}\mathrm{\mathbf{v}}_{a}+k_b\mathrm{\mathbf{v}}_{b} - \mathrm{\mathbf{u}}_r \right] \:\mathrm{mod}\:\Lambda \right)
\label{input-relay-first-strategy}
\end{equation}
during the second transmission period, where the scalar $\beta_r$ is chosen so as to adjust its transmitted power during this period.

\textit{Decoding and Rate Analysis:}

\underline{\textit{Step 1)}} During the first transmission period, the relay receives
\begin{eqnarray}
\mathrm{\mathbf{y}}_{r} &=& h_{ar}\mathrm{\mathbf{x}}_{a} + h_{br}\mathrm{\mathbf{x}}_{b}+ \mathrm{\mathbf{z}}_{r}.
\label{output-relay-first-transmission-period-first-strategy}
\end{eqnarray}

\noindent Next, the relay performs the following modulo-reduction operation on the received signal:
\begin{align}
\mathrm{\mathbf{y}}_{r}' &= [\alpha_r\mathrm{\mathbf{y}}_{r}+ k_{a}\mathrm{\mathbf{u}}_{a}+ k_b\mathrm{\mathbf{u}}_{b}]\:\mathrm{mod}\:\Lambda \nonumber\\
&= \big[\alpha_r(h_{ar}\dv x_a+h_{br}\dv x_b+\dv z_r)+ \frac{k_a}{\beta_a}\dv x_a- \frac{k_a}{\beta_a}\dv x_a + \frac{k_b}{\beta_b}\dv x_b- \frac{k_b}{\beta_b}\dv x_b + k_{a}\mathrm{\mathbf{u}}_{a}+ k_b\mathrm{\mathbf{u}}_{b}\big]\:\mathrm{mod}\:\Lambda \nonumber\\
&\stackrel{(a)}{=} \big[(\alpha_rh_{ar}-\frac{k_a}{\beta_a})\dv x_a+(\alpha_rh_{br}-\frac{k_b}{\beta_b})\dv x_b + k_a[\dv v_a-\dv u_a]\:\mathrm{mod}\:\Lambda + k_b[\dv v_b-\dv u_b]\:\mathrm{mod}\:\Lambda + \alpha_r\dv z_r + k_{a}\mathrm{\mathbf{u}}_{a}+ k_b\mathrm{\mathbf{u}}_{b}\big]\:\mathrm{mod}\:\Lambda \nonumber\\
&\stackrel{(b)}{=} \big[k_{a}\mathrm{\mathbf{v}}_{a}+k_b\mathrm{\mathbf{v}}_{b}+(\alpha_rh_{ar}-\frac{k_a}{\beta_a})\dv x_a+(\alpha_rh_{br}-\frac{k_b}{\beta_b})\dv x_b + \alpha_r\dv z_r\big]\:\mathrm{mod}\:\Lambda \nonumber\\
&\stackrel{(c)}{=} [k_{a}\mathrm{\mathbf{v}}_{a}+k_b\mathrm{\mathbf{v}}_{b}+\mathrm{\mathbf{z}}_{r}']\:\mathrm{mod}\:\Lambda
\label{equivalent-output-relay-first-transmission-period-first-strategy}
\end{align}
where $(a)$ follows by substituting $\dv x_a$ and  $\dv x_b$ using \eqref{input-first-transmitter-first-strategy} and \eqref{input-second-transmitter-first-strategy}; $(b)$ follows by using the properties (P1) and (P2) in \eqref{properties-modulo-operation}; and $(c)$ follows by substituting,
\begin{equation}
\mathrm{\mathbf{z}}_{r}' \triangleq \big[\alpha_r\dv z_r+(\alpha_rh_{ar}-\frac{k_a}{\beta_a})\dv x_a+(\alpha_rh_{br}-\frac{k_b}{\beta_b})\dv x_b \big]\:\mathrm{mod}\:\Lambda.
\label{effective-noise-relay-first-transmission-period-first-strategy}
\end{equation}
 
\noindent The parameter $\alpha_r \in \mathbb{R}$ is some inflation factor whose optimal value will be specified below and $\mathrm{\mathbf{z}}_{r}'$ is the \textit{effective} noise at the relay. Since an integer combination of lattice points is a lattice point, $\dv e_2=[k_{a}\mathrm{\mathbf{v}}_{a}+k_b\mathrm{\mathbf{v}}_{b}] \in \Lambda$ and, thus, the equivalent channel $\mathrm{\mathbf{y}}_{r}'$ is a modulo-lattice additive noise (MLAN) channel \cite{F88} with noise equal to $\mathrm{\mathbf{z}}_{r}'$. Hence, the probability of error $\text{Pr}(\hat{\dv e}_2 \ne \dv e_2)$ is equal to the probability that the equivalent noise leaves the Voronoi region surrounding the codeword, i.e., $\text{Pr}(\mathrm{\mathbf{z}}_{r}' \notin \mc V_{\text{FINE}})$ \cite{NG11a}. As we will show shortly this probability can be made as small as desired; and, thus, the relay obtains the integer combination of the users' codewords $\dv e_2=[k_{a}\mathrm{\mathbf{v}}_{a}+k_b\mathrm{\mathbf{v}}_{b}]$ from the MLAN channel $\mathrm{\mathbf{y}}_{r}'$ correctly. More specifically, consider the channel from $\dv e_2$ to $\dv y'_r$ as given by \eqref{equivalent-output-relay-first-transmission-period-first-strategy}. Due to the dither $\dv u_a$, the input $\dv x_a$ of Transmitter $A$ is independent of $\dv v_a=\phi_a(W_a) \in \mc C^{}$, and is uniformly distributed over $\mc V$ (see, e.g., \cite{F03,ESZ05}). Similarly, due to the dither $\dv u_b$, the input $\dv x_b$ of Transmitter $B$ is independent of $\dv v_b=\phi_b(W_b) \in \mc C^{}$, and is uniformly distributed over $\mc V$. The effective noise $\dv z'_r$ is then independent of $\dv v_a$ and $\dv v_b$, and so of $\dv e_2 = k_a\dv v_a + k_b\dv v_b$. It is composed of a ``self noise" component and a Gaussian noise component. Proceeding essentially as in \cite{NG11a}, the density of $\dv z_r'$ can be upper bounded by the density of an i.i.d. zero-mean Gaussian vector $\dv z_r^*$ whose variance approaches
  \begin{equation}
  \text{Var}\:(\mathrm{\mathbf{z}}^{\star}_{r}) = \alpha_r^2N+P(\alpha_rh_{ar}\beta_a-k_a)^2+P(\alpha_rh_{br}\beta_b-k_b)^2
  \label{bound-variance-effective-noise-relay-first-transmission-period-first-strategy}
  \end{equation}
  as $n \longrightarrow \infty$.

\noindent Since the lattice $\Lambda_{\text{FINE}}$ is chosen to be good for AWGN, the probability that noise $\dv z^{\star}_r$ leaves the Voronoi region $\mc V_{\text{FINE}}$ of lattice $\Lambda_{\text{FINE}}$ goes to zero exponentially in $n$ as long as
\begin{equation}
\text{Vol}(\mc V_{\text{FINE}}) > \Big(2\pi{e}\text{Var}\:(\mathrm{\mathbf{z}}^{\star}_{r})\Big)^{n/2}. 
\end{equation}
If this occurs, $\text{Pr}(\dv z'_r \notin \mc V_{\text{FINE}})$ goes to zero exponentially in $n$ as well. Noticing that the variance of the Gaussian noise $\dv z_r^*$ depends on the choice of the inflation parameter $\alpha_r$, the probability of error $\text{Pr}(\dv z'_r \notin \mc V_{\text{FINE}})$ of course also goes to zero exponentially in $n$ if we set the Voronoi region $\mc V_{\text{FINE}}$ of lattice $\Lambda_{\text{FINE}}$ to satisfy the constraint
\begin{equation}
\text{Vol}(\mc V_{\text{FINE}}) > \Big(2\pi{e} \max_{\alpha_r} \text{Var}\:(\mathrm{\mathbf{z}}^{\star}_{r})\Big)^{n/2}. 
\label{condition-volume-decoding-at-relay}
\end{equation}

\noindent The right hand side (RHS) of \eqref{condition-volume-decoding-at-relay} is maximized by setting the inflation parameter $\alpha_r$ to 
\begin{equation}
\alpha_r^{\star} = \frac{P(\boldsymbol{\beta_s} \circ \mathrm{\mathbf{h}}_r)^T\mathrm{\mathbf{k}}}{N+P\|\boldsymbol{\beta_s} \circ \mathrm{\mathbf{h}}_r\|^2}.
\label{optimal-choice-inflation-parameter-decoding-at-relay-first-strategy}
\end{equation}

\noindent Recall that the rate of the nested code that we employ is
\begin{equation}
R=\frac{1}{2n}\log\Big(\frac{\text{Vol}(\mc V)}{\text{Vol}(\mc V_{\text{FINE}})}\Big).
\end{equation}
Solving for the volume of the fine lattice $\Lambda_{\text{FINE}}$, and then substituting using \eqref{bound-variance-effective-noise-relay-first-transmission-period-first-strategy}, \eqref{condition-volume-decoding-at-relay} and \eqref{optimal-choice-inflation-parameter-decoding-at-relay-first-strategy}, we get that $\text{Pr}(\dv z'_r \notin \mc V_{\text{FINE}})$ goes to zero exponentially in $n$ if
\begin{align}
R &< \frac{1}{4}\log^{+}\left(\left(\|\mathrm{\mathbf{k}}\|^2-\frac{P((\boldsymbol{\beta_s} \circ \mathrm{\mathbf{h}}_r)^T \mathrm{\mathbf{k}})^2}{N+P\|\boldsymbol{\beta_s} \circ \mathrm{\mathbf{h}}_r\|^2}\right)^{-1}\right) - \frac{1}{4}\log\left(2\pi e G(\Lambda)\right).
\label{computation-rate-relay-first-transmission-period-first-strategy}
\end{align}

\underline{\textit{Step 2)}} During the first transmission period, the destination receives
\begin{equation}
\mathrm{\mathbf{y}}_{d} = h_{ad}\mathrm{\mathbf{x}}_{a} + h_{bd}\mathrm{\mathbf{x}}_{b}+ \mathrm{\mathbf{z}}_{d}.
\label{output-destination-first-transmission-period-first-strategy}
\end{equation}
Similar to the relay, the destination computes a linear combination with integer coefficients of the transmitters' codewords by performing the modulo-reduction operation:
\begin{align}
\mathrm{\mathbf{y}}_{d}' &= [\alpha_d\mathrm{\mathbf{y}}_{d}+ t_{a}\mathrm{\mathbf{u}}_{a}+ t_b\mathrm{\mathbf{u}}_{b}]\:\mathrm{mod}\:\Lambda \nonumber\\
&= [t_{a}\mathrm{\mathbf{v}}_{a}+t_b\mathrm{\mathbf{v}}_{b}+\mathrm{\mathbf{z}}_{d}']\:\mathrm{mod}\:\Lambda
\label{equivalent-output-destination-first-transmission-period-first-strategy}
\end{align}
where $\alpha_d \in \mathbb{R}$ is some inflation factor and $\mathrm{\mathbf{z}}_{d}'$ is the effective noise at the destination, given by
\begin{equation}
\mathrm{\mathbf{z}}_{d}' \triangleq \big[\alpha_d\dv z_d+(\alpha_dh_{ad}-\frac{t_a}{\beta_a})\dv x_a+(\alpha_dh_{bd}-\frac{t_b}{\beta_b})\dv x_b \big]\:\mathrm{mod}\:\Lambda.
\label{effective-noise-destination-first-transmission-period-first-strategy}
\end{equation}
Thus, during the first transmission period, by using the MLAN channel $\mathrm{\mathbf{y}}_{d}'$, the destination can obtain a first integer combination $\dv e_1=[t_{a}\mathrm{\mathbf{v}}_{a}+t_b\mathrm{\mathbf{v}}_{b}]$ of the users' codewords using its output component from the direct links. The probability of error incurred during this step is equal to the probability that the equivalent noise $\mathrm{\mathbf{z}}_{d}'$ leaves the Voronoi region surrounding the codeword, i.e., $\text{Pr}(\mathrm{\mathbf{z}}_{d}' \notin \mc V_{\text{FINE}})$. Using analysis and algebra that are essentially similar to those for decoding at the relay node, this probability of error can be shown to go to zero exponentially in $n$ if
\begin{align}
R &<  \frac{1}{4}\log^{+}\left(\left(\|\mathrm{\mathbf{t}}\|^2-\frac{P((\boldsymbol{\beta_s} \circ \mathrm{\mathbf{h}}_d)^T \mathrm{\mathbf{t}})^2}{N+P\|\boldsymbol{\beta_s} \circ \mathrm{\mathbf{h}}_d\|^2}\right)^{-1}\right) - \frac{1}{4}\log\left(2\pi e G(\Lambda)\right).
\label{computation-rate-destination-first-transmission-period-first-strategy}
\end{align}

\underline{\textit{Step 3)}} During the second transmission period, the destination receives from the relay the signal,
\begin{align}
\mathrm{\mathbf{\tilde{y}}}_{d} &= h_{rd}\mathrm{\mathbf{\tilde{x}}}_{r} + \mathrm{\mathbf{\tilde{z}}}_{d}\nonumber\\
&=  h_{rd}\beta_r\left(\left[k_{a}\mathrm{\mathbf{v}}_{a}+k_b\mathrm{\mathbf{v}}_{b} - \mathrm{\mathbf{u}}_r \right] \:\mathrm{mod}\:\Lambda \right) + \mathrm{\mathbf{\tilde{z}}}_{d}.
\label{output-destination-second-transmission-period-first-strategy}
\end{align}

\noindent Again, by performing a modulo-reduction operation on the obtained signal, the destination gets
\begin{align}
\mathrm{\mathbf{\tilde{y}}}_{d}' &= [\tilde{\alpha}_d\mathrm{\mathbf{\tilde{y}}}_{d} + \mathrm{\mathbf{u}}_{r}]\:\mathrm{mod}\:\Lambda \nonumber\\
&= [k_{a}\mathrm{\mathbf{v}}_{a}+k_b\mathrm{\mathbf{v}}_{b}+ \mathrm{\mathbf{\tilde{z}}}_{d}']\:\mathrm{mod}\:\Lambda,
\label{equivalent-output-destination-second-transmission-period-first-strategy}
\end{align} 
where $\tilde{\alpha}_d \in \mathbb{R}$ is some inflation factor and $\mathrm{\mathbf{\tilde{z}}}_{d}'$ is the effective noise at the destination, given by
\begin{equation}
\mathrm{\mathbf{\tilde{z}}}_{d}' \triangleq \big[\tilde{\alpha}_d\mathrm{\mathbf{\tilde{z}}}_{d} + (\tilde{\alpha}_dh_{rd}-\frac{1}{\beta_r})\mathrm{\mathbf{\tilde{x}}}_{r}\big]\:\mathrm{mod}\:\Lambda.
\label{effective-noise-destination-second-transmission-period-first-strategy}
\end{equation}
Thus, during the second transmission period, the destination can obtain a second integer combination $\dv e_2=[k_{a}\mathrm{\mathbf{v}}_{a}+k_b\mathrm{\mathbf{v}}_{b}]$ of the users' codewords using its output component from the relay. (Recall that this combination has first been computed at the relay at the end of the first transmission period, and then forwarded to the destination during the second transmission period). Again, proceeding in a way that is similar to above, it can be shown that this can be accomplished with small probability of error $\text{Pr}(\mathrm{\mathbf{\tilde{z}}}_{d}' \notin \mc V_{\text{FINE}})$ if 
\begin{equation}
R < \frac{1}{4}\log\left(1+\frac{P|h_{rd}|^2\beta^2_r}{N}\right)- \frac{1}{4}\log\left(2\pi e G(\Lambda)\right).
\label{computation-rate-destination-second-transmission-period-first-strategy}
\end{equation}

\underline{\textit{Summarizing:}} The probability of error goes to zero for all desired equations if \eqref{computation-rate-relay-first-transmission-period-first-strategy}, \eqref{computation-rate-destination-first-transmission-period-first-strategy} and \eqref{computation-rate-destination-second-transmission-period-first-strategy} hold simultaneously. If this holds, over the entire transmission time, the destination collects two linear combinations with integer coefficients that relate the users' codewords, as
\begin{eqnarray}
\left[ \begin{array}{ccc}
\dv e_1\\
\dv e_2
\end{array} \right] = \left[ \begin{array}{ccc}
t_{a} & t_{b}  \\
k_{a} & k_{b}  \\
\end{array} \right]
\left[ \begin{array}{ccc}
\mathrm{\mathbf{v}}_{a}\\
\mathrm{\mathbf{v}}_{b}\\
\end{array} \right].
\label{system-of-equalities-at-destination-first-strategy}
\end{eqnarray}
Now, since the integer-valued matrix in \eqref{system-of-equalities-at-destination-first-strategy} is invertible (recall that the integer-valued coefficients are chosen such that $\text{det}(\dv k, \dv t) \neq  0$), the destination obtains the transmitted codewords by solving \eqref{system-of-equalities-at-destination-first-strategy}.

\noindent The destination is able to recover the messages $\hat{W}_a$ and $\hat{W}_b$ reliably if the message rate is less or equal to the computational rate \cite{NG11a}. Then, let us define $R_{\text{sr}}(\Lambda)$ as the RHS of \eqref{computation-rate-relay-first-transmission-period-first-strategy}, $R_{\text{sd}}(\Lambda)$ as the RHS of \eqref{computation-rate-destination-first-transmission-period-first-strategy}, and $R_{\text{rd}}(\Lambda)$ as the RHS of \eqref{computation-rate-destination-second-transmission-period-first-strategy}. The above means that using the coding scheme that we described, which employs the $n$-dimensional lattice $\Lambda$, the destination can decode the transmitters' codewords correctly at a transmission rate that is equal to the minimum among $R_{\text{sr}}(\Lambda)$, $R_{\text{sd}}(\Lambda)$ and 
 $R_{\text{rd}}(\Lambda)$, i.e., $R(\Lambda) = \min\{R_{\text{sr}}(\Lambda), R_{\text{sd}}(\Lambda), R_{\text{rd}}(\Lambda)\}$. The allowed symmetric-rate is given by $R^{\text{CoF}}_{\text{sym}}(\Lambda)=R(\Lambda)=\min\{R_{\text{sr}}(\Lambda), R_{\text{sd}}(\Lambda), R_{\text{rd}}(\Lambda)\}$. Noticing that $2\pi e G(\Lambda) \to 1$ when $n \to \infty$ \cite{CS88}, the desired symmetric-rate \eqref{achievable-sum-rate-compute-and-forward-at-relay} is obtained by taking the limit of $R^{\text{CoF}}_{\text{sym}}(\Lambda)$ as $n$ goes to infinity; and this completes the proof of Proposition~\ref{proposition-achievable-sum-rate-compute-and-forward-at-relay}. \qed

\vspace{0.3cm}

\begin{remark}
The scheme of Proposition~\ref{proposition-achievable-sum-rate-compute-and-forward-at-relay} is conceptually similar to the compute-and-forward approach of Nazer and Gastpar \cite{NG11a}. This can be seen by noticing that the multiaccess relay network that we study in this paper can be thought as being a Gaussian network with two users, two relays and a central processor. The first relay in the equivalent network plays the role of the relay in our MARC model, and the second relay in the equivalent network plays the role of the destination in our MARC model. The second relay in the equivalent network is connected with the central processor, which is the destination itself, via a bit-pipe of infinite capacity. Furthermore, it can be seen that, in the equivalent model, the bit-pipe with infinite capacity can be replaced with one that has the same capacity as that of the relay-to-destination link. This follows since the two equations that are forwarded to the central processor have the same rate. Hence, the rate of Proposition~\ref{proposition-achievable-sum-rate-compute-and-forward-at-relay} can also be readily obtained by viewing the MARC network that we study as described in this remark and then applying the result of \cite[Theorem 5]{NG11a}.
\end{remark}

\begin{remark}\label{binning}
As we mentioned previously, in the coding scheme of Proposition~\ref{proposition-achievable-sum-rate-compute-and-forward-at-relay} the relay decodes a linear combination $\dv e_2$ of the users' codewords and then sends it to the destination during the second transmission period. Noticing that the destination already observes side information $\dv e_1$ that is computed from the direct transmission from the sources, the rate of Proposition~\ref{proposition-achievable-sum-rate-compute-and-forward-at-relay} can be improved by having the relay convey the decoded equation $\dv e_2$ to the destination during the second transmission period losslessly using Slepian-Wolf binning. This increases the transmission rate by reducing the rate cost that is associated with conveying the decoded equation $\dv e_2$ to the destination.
\end{remark}

Although the achievable symmetric-rate in Proposition~\ref{proposition-achievable-sum-rate-compute-and-forward-at-relay} requires the relay to only decode a linear combination of the codewords transmitted by the users, not the individual messages, this can be rather a severe constraint in certain cases. In the following section, the relay only compresses its output and sends it to the destination. The computation of the desired linear combinations of the users' codewords takes place at the destination, locally.  


\subsection{Compress-and-Forward at the Relay and Compute at the Destination}\label{secIII_subsecB}

\vspace{0.1cm}

The following proposition provides an achievable symmetric-rate for the multiaccess relay model that we study.

\vspace{0.2cm}

\begin{proposition}\label{proposition-achievable-sum-rate-compress-and-forward-at-relay-and-compute-at-destination}
For any set of channel vector $\mathrm{\mathbf{h}} = [h_{ar},\: h_{br},\: h_{ad},\: h_{bd},\: h_{rd}]^T \in \mathbb{R}^5$, the following symmetric-rate is achievable:
\begin{align}
R^{\text{CoD}}_{\text{sym}} = \max \: \frac{1}{4}\min\Bigg\{&\log^{+}\left(\frac{\text{snr}}{\text{snr}||\boldsymbol{\beta}_s \circ \dv H^T\boldsymbol{\alpha}_t -\dv t||^2+(\boldsymbol{\alpha}_t \circ \boldsymbol{\alpha}_t)^T \dv n_d}\right),\nonumber\\
& \log^{+}\left(\frac{\text{snr}}{\text{snr}||\boldsymbol{\beta}_s \circ \dv H^T\boldsymbol{\alpha}_k-\dv k||^2 +(\boldsymbol{\alpha}_k \circ \boldsymbol{\alpha}_k)^T \dv n_d}\right)\Bigg\},
\label{achievable-sum-rate-compress-and-forward-at-relay-and-compute-at-destination}
\end{align}
where $\boldsymbol{\alpha}_t= [\alpha_{1t},\: \alpha_{2t}]^T$ and $\boldsymbol{\alpha}_k= [\alpha_{1k},\: \alpha_{2k}]^T$ $\in \mathbb{R}^{2}$ are some inflation factors, $\dv n_d=[1,\: 1+D/N]^T$ $\in \mathbb{R}^{2}$, and $D$ is given by
\begin{equation}
D = \frac{N^2\left(1+\text{snr}\|\boldsymbol{\beta_s} \circ \dv h_r\|^2\right)}{|h_{rd}|^2P_r}-\frac{N^2\big(\text{snr}(\boldsymbol{\beta_s}\circ\dv h_r)^T(\boldsymbol{\beta_s}\circ\dv h_d)\big)^2}{|h_{rd}|^2P_r\left(1+\text{snr}\|\boldsymbol{\beta_s} \circ \dv h_d\|^2\right)}, 
\end{equation}
and the maximization is over $\boldsymbol{\alpha}_t$, $\boldsymbol{\alpha}_k$, $\boldsymbol{\beta_s}$, and over the integer coefficients $\dv k$ and $\dv t$ such that $|\text{det}(\dv k, \dv t)| \geq 1$.
\end{proposition}

\vspace{0.2cm}

As we indicated previously, in the coding scheme that we use for the proof of Proposition~\ref{proposition-achievable-sum-rate-compress-and-forward-at-relay-and-compute-at-destination}, the relay conveys a lossy version of its output to the destination during the second transmission period. In doing so, it accounts for the available side information at the destination, i.e, what the destination has received during the first transmission period. The destination computes two linearly independent combinations that relate the users' codewords using its outputs from both transmission periods, as follows. The destination combines appropriately the obtained lossy version of the relay's output (that it recovered from the relay's transmission during the second transmission period) and from what it received during the first transmission period. Then it computes two linearly independent combinations with integer coefficients that relate the users' codewords. 

\vspace{0.2cm}

\textbf{Proof:} The transmission scheme, and the encoding procedures at the transmitters are similar to those in the proof of Proposition~\ref{proposition-achievable-sum-rate-compute-and-forward-at-relay}. Therefore they will be outlined only, for brevity. We will insist more on aspects of the coding scheme that are inherently different from those of the coding scheme of Proposition~\ref{proposition-achievable-sum-rate-compute-and-forward-at-relay}. 


\textit{Encoding:} During the first transmission period, the transmitters send the same inputs as in the coding scheme of Proposition~\ref{proposition-achievable-sum-rate-compute-and-forward-at-relay}, i.e., to transmit message $W_a$, Transmitter $A$ sends the input $\dv x_a$ given by \eqref{input-first-transmitter-first-strategy}; and to transmit message $W_b$, Transmitter $B$ sends the input $\dv x_b$ given by \eqref{input-second-transmitter-first-strategy}.

The relay quantizes what it receives during the first transmission period using Wyner-Ziv compression \cite{WZ76}, accounting for the available side information $\dv y_d$ at the destination. Let $\mathrm{\mathbf{\hat{y}}}_r$ be the compressed version of $\dv y_r$ given by
\begin{align}
\mathrm{\mathbf{\hat{y}}}_r &= \dv y_r + \dv d                             
\label{compressed-version-relay-output-second-strategy}
\end{align}
where $\dv d$ is a Gaussian random vector whose elements are i.i.d with zero mean and variance $D$; and is independent of all other signals. Also let $\hat{R}_{\text{WZ}}$ be the resulting compression rate. During the second transmission period, the relay conveys the description $\mathrm{\mathbf{\hat{y}}}_r$ of $\dv y_r$ to the destination. To this end, it sends an independent Gaussian input $\mathrm{\mathbf{\tilde{x}}}_{r}$ with $\beta^2_rP$ and carries the Wyner-Ziv compression index of $\mathrm{\mathbf{\hat{y}}}_r$.

\textit{Decoding at the destination:} During the two transmission periods, the destination receives,
\begin{eqnarray}
\mathrm{\mathbf{y}}_{d} &=& h_{ad}\mathrm{\mathbf{x}}_{a} + h_{bd}\mathrm{\mathbf{x}}_{b}+ \mathrm{\mathbf{z}}_{d}\nonumber \\
\mathrm{\mathbf{\tilde{y}}}_{d} &=& h_{rd}\mathrm{\mathbf{\tilde{x}}}_{r}+\mathrm{\mathbf{\tilde{z}}}_{d}.
\label{outputs-destination-second-strategy}
\end{eqnarray}

The destination computes two linearly independent combinations with integer coefficients that relate the users' codewords, as follows.

\underline{\textit{Step 1)}} It first recovers the compressed version of the relay's output sent by the relay during the second transmission period, by utilizing its output $\mathrm{\mathbf{\tilde{y}}}_{d}$ as well as the available side information $\mathrm{\mathbf{y}}_{d}$. As it will be shown below, the destination recovers the compressed version $\mathrm{\mathbf{\hat{y}}}_r$ of $\dv y_r$ if the constraint \eqref{constraint-distortion-second-strategy} below is satisfied (see the ``Rate Analysis" section).

\noindent Next, the destination combines $\mathrm{\mathbf{y}}_{d}$ and $\mathrm{\mathbf{\hat{y}}}_r$ and uses the obtained signal to compute a linear combination with integer coefficients of the users' codewords \cite{ZNEG10}. More specifically, let  
\begin{align}
\mathrm{\mathbf{y}}_t &= \alpha_{1t}\mathrm{\mathbf{y}}_{d} + \alpha_{2t}\mathrm{\mathbf{\hat{y}}}_r \nonumber\\
                    &= \big(\alpha_{1t}h_{ad}+\alpha_{2t}h_{ar}\big)\dv x_a + \big(\alpha_{1t}h_{bd}+\alpha_{2t}h_{br}\big)\dv x_b + \alpha_{1t} \dv z_d + \alpha_{2t} \dv z_r + \alpha_{2t} \dv d,
\label{maximum-ratio-combined-output-destination-second-strategy1}
\end{align} 
for some $\boldsymbol{\alpha}_t = [\alpha_{1t},\alpha_{2t}]^T \in \mathbb{R}^2$. The destination uses the obtained signal $\dv y_t$ to compute a linear combination with integer coefficients of the transmitters' codewords by performing the modulo reduction operation
\begin{align}
\mathrm{\mathbf{y}}_t' &= [\mathrm{\mathbf{y}}_t + t_{a}\mathrm{\mathbf{u}}_{a} + t_b\mathrm{\mathbf{u}}_{b}]\:\mathrm{mod}\:\Lambda \nonumber\\
&= [t_{a}\mathrm{\mathbf{v}}_{a}+t_b\mathrm{\mathbf{v}}_{b}+\mathrm{\mathbf{z}}_t']\:\mathrm{mod}\:\Lambda
\label{equivalent-maximum-ratio-combined-output-destination-second-strategy1}
\end{align}
where the algebra follows \eqref{equivalent-output-relay-first-transmission-period-first-strategy} and $\mathrm{\mathbf{z}}_t'$ is the effective noise given by
\begin{align}
\mathrm{\mathbf{z}}_t' &\triangleq \Big[\alpha_{1t}\dv z_d + \alpha_{2t}\dv z_r + \alpha_{2t}\dv d+(\alpha_{1t}h_{ad}+\alpha_{2t}h_{ar}-\frac{t_a}{\beta_a})\dv x_a+(\alpha_{1t}h_{bd}+\alpha_{2t}h_{br}-\frac{t_b}{\beta_b})\dv x_b \Big]\:\mathrm{mod}\:\Lambda.
\label{effective-noise-maximum-ratio-combined-output-destination-second-strategy1}
\end{align}
Finally, by decoding the lattice point $\dv e_1 = [t_a\dv v_a + t_b\dv v_b] \in \Lambda$ using the MLAN channel $\mathrm{\mathbf{y}}_t'$, the destination obtains a first linear combination with integer coefficients of the users' codewords. As it will be shown below, this can be accomplished with a probability of error $\text{Pr}(\mathrm{\mathbf{z}}_{t}' \notin \mc V_{\text{FINE}})$ that is as small as desired. 

\underline{\textit{Step 2}} Second, the destination again combines $\mathrm{\mathbf{y}}_{d}$ and $\mathrm{\mathbf{\hat{y}}}_r$ and uses the obtained signal to compute a second linear combination with integer coefficients of the users' codewords which is different from the one decoded in \textit{Step 1}. More specifically, let  
\begin{align}
\mathrm{\mathbf{y}}_k &= \alpha_{1k}\mathrm{\mathbf{y}}_{d} + \alpha_{2k}\mathrm{\mathbf{\hat{y}}}_r \nonumber\\
                    &= \big(\alpha_{1k}h_{ad}+\alpha_{2k}h_{ar}\big)\dv x_a + \big(\alpha_{1k}h_{bd}+\alpha_{2k}h_{br}\big)\dv x_b + \alpha_{1k} \dv z_d + \alpha_{2k} \dv z_r + \alpha_{2k} \dv d,
\label{maximum-ratio-combined-output-destination-second-strategy2}
\end{align} 
for some $\boldsymbol{\alpha}_k = [\alpha_{1k},\alpha_{2k}]^T \in \mathbb{R}^2$. The destination uses the obtained signal $\dv y_k$ to compute a linear combination with integer coefficients of the transmitters' codewords by performing the modulo reduction operation
\begin{align}
\mathrm{\mathbf{y}}_k' &= [\mathrm{\mathbf{y}}_k + k_{a}\mathrm{\mathbf{u}}_{a} + k_b\mathrm{\mathbf{u}}_{b}]\:\mathrm{mod}\:\Lambda \nonumber\\
&= [k_{a}\mathrm{\mathbf{v}}_{a}+k_b\mathrm{\mathbf{v}}_{b}+\mathrm{\mathbf{z}}_k']\:\mathrm{mod}\:\Lambda
\label{equivalent-maximum-ratio-combined-output-destination-second-strategy2}
\end{align}
where the algebra follows \eqref{equivalent-output-relay-first-transmission-period-first-strategy} and $\mathrm{\mathbf{z}}_k'$ is the effective noise given by
\begin{align}
\mathrm{\mathbf{z}}_k' &\triangleq \Big[\alpha_{1k}\dv z_d + \alpha_{2k}\dv z_r + \alpha_{2k}\dv d+(\alpha_{1k}h_{ad}+\alpha_{2k}h_{ar}-\frac{k_a}{\beta_a})\dv x_a+(\alpha_{1k}h_{bd}+\alpha_{2k}h_{br}-\frac{k_b}{\beta_b})\dv x_b \Big]\:\mathrm{mod}\:\Lambda.
\label{effective-noise-maximum-ratio-combined-output-destination-second-strategy2}
\end{align}
Finally, by decoding the lattice point $\dv e_2 = [k_a\dv v_a + k_b\dv v_b] \in \Lambda$ using the MLAN channel $\mathrm{\mathbf{y}}_k'$, the destination obtains a second linear combination with integer coefficients of the users' codewords. This can be accomplished with a probability of error $\text{Pr}(\mathrm{\mathbf{z}}_{k}' \notin \mc V_{\text{FINE}})$ that is as small as desired.

\textit{Rate Analysis:} 

The relay compresses its output $\dv y_r$ at the per-channel use Wyner-Ziv quantization rate \cite{WZ76}
\begin{align}
\hat{R}_{\text{WZ}}  &= \frac{1}{2n}I(\mathrm{\mathbf{y}}_r;\mathrm{\mathbf{\hat{y}}}_{r}|\mathrm{\mathbf{y}}_{d})\nonumber\\ 
&= \frac{1}{2n}h(\mathrm{\mathbf{\hat{y}}}_{r}|\mathrm{\mathbf{y}}_{d})-\frac{1}{2n}h(\mathrm{\mathbf{\hat{y}}}_{r}|\mathrm{\mathbf{y}}_r,\mathrm{\mathbf{y}}_{d})\nonumber\\
&\stackrel{(a)}{=} \frac{1}{2n}h(\mathrm{\mathbf{\hat{y}}}_{r}|\mathrm{\mathbf{y}}_{d})-\frac{1}{2n}h(\mathrm{\mathbf{d}})\nonumber\\
&\stackrel{(b)}{\leq}  \frac{1}{4n}\log(2{\pi}e)^n\Big|\mathbb{E}[\mathrm{\mathbf{\hat{y}}}_{r}\mathrm{\mathbf{\hat{y}}}_{r}^T]-\mathbb{E}[\mathrm{\mathbf{\hat{y}}}_{r}\mathbb{E}[\mathrm{\mathbf{\hat{y}}}_{r}|\mathrm{\mathbf{y}}_{d}]^T]\Big|-\frac{1}{4}\log\left(2 \pi e D\right)\nonumber\\
&\stackrel{(c)}{=}  \frac{1}{4}\log\left(1+\frac{N+P\|\boldsymbol{\beta}_s \circ \mathrm{\mathbf{h}}_r\|^2}{D}-\frac{[P(\boldsymbol{\beta}_s\circ\mathrm{\mathbf{h}}_r)^T(\boldsymbol{\beta}_s\circ\mathrm{\mathbf{h}}_d)]^2}{D(N+P\|\boldsymbol{\beta}_s \circ \mathrm{\mathbf{h}}_d\|^2)}\right),
\label{Wyner-Ziv-quantization-rate-second-strategy}
\end{align}
where $(a)$ follows since $\dv d$ is independent of $(\dv y_r,\dv y_d)$; $(b)$ follows since, by the \textit{maximum conditional differential entropy lemma} \cite[Chapter 2, p. 21]{GK11}, the conditional entropy $h(\mathrm{\mathbf{\hat{y}}}_{r}|\mathrm{\mathbf{y}}_{d})$ is upper-bounded by that of jointly Gaussian signals of the same covariance matrix, and $\dv d$ is an i.i.d Gaussian vector; $(c)$ follows through by straightforward algebra, and by noticing that the minimum mean square error (MMSE) estimator of $\mathrm{\mathbf{\hat{y}}}_{r}$ given $\mathrm{\mathbf{y}}_{d}$ is given by
\begin{align}
\mathbb{E}[\mathrm{\mathbf{\hat{y}}}_{r}|\mathrm{\mathbf{y}}_{d}] &= \frac{P(\boldsymbol{\beta}_s\circ\mathrm{\mathbf{h}}_r)^T(\boldsymbol{\beta}_s\circ\mathrm{\mathbf{h}}_d)}{N+P\|\boldsymbol{\beta}_s \circ \mathrm{\mathbf{h}}_d\|^2} \: \mathrm{\mathbf{y}}_{d}.
\end{align}


At the end of the second transmission period, the destination can decode the correct relay input $\mathrm{\mathbf{\tilde{x}}}_{r}$ reliably if
\begin{eqnarray}
\hat{R}_{\text{WZ}}  &\leq& \frac{1}{4}\log\left(1+\frac{P|h_{rd}|^2\beta_r^2}{N}\right).
\label{constraint-source-channel-separation-decoding-at-destination-second-strategy}
\end{eqnarray}
From \eqref{Wyner-Ziv-quantization-rate-second-strategy} and \eqref{constraint-source-channel-separation-decoding-at-destination-second-strategy}, we get the following constraint on the distortion 
\begin{align}
D &\geq \frac{N^2\left(1+\text{snr}\|\boldsymbol{\beta_s} \circ \dv h_r\|^2\right)}{|h_{rd}|^2P_r}-\frac{N^2\big(\text{snr}(\boldsymbol{\beta_s}\circ\dv h_r)^T(\boldsymbol{\beta_s}\circ\dv h_d)\big)^2}{|h_{rd}|^2P_r\left(1+\text{snr}\|\boldsymbol{\beta_s} \circ \dv h_d\|^2\right)}. 
\label{constraint-distortion-second-strategy}
\end{align}
The above implies that, under the constraint \eqref{constraint-distortion-second-strategy}, the destination recovers the lossy version $\mathrm{\mathbf{\hat{y}}}_r$ of what was sent by the relay during the second transmission period.

The destination processes $\mathrm{\mathbf{y}}_t$ and $\mathrm{\mathbf{y}}_k$ to obtain the linear combinations $\dv e_1 = [t_a\dv v_a + t_b\dv v_b]$ , and $\dv e_2 = [k_a\dv v_a + k_b\dv v_b]$ of the users' codewords. 

Using the MLAN channel $\dv y_t'$ given by \eqref{equivalent-maximum-ratio-combined-output-destination-second-strategy1} and proceeding in a way that is essentially similar to in the proof of Proposition~\ref{proposition-achievable-sum-rate-compute-and-forward-at-relay}, it can be shown that in decoding the linear combination $\dv e_1 = t_a\dv v_a + t_b\dv v_b$, the probability of error at the destination $\text{Pr}(\mathrm{\mathbf{z}}_{t}' \notin \mc V_{\text{FINE}})$ goes to zero exponentially in $n$ if 
\begin{align}
R &< \:\frac{1}{4}\:\log^{+}\left(\frac{\text{snr}}{\text{snr}||\boldsymbol{\beta}_s \circ \dv H^T\boldsymbol{\alpha}_t -\dv t||^2+(\boldsymbol{\alpha}_t \circ \boldsymbol{\alpha}_t)^T \dv n_d}\right)-\frac{1}{4}\log\left(2\pi e G(\Lambda)\right),
\label{first-computation-rate-destination-second-strategy}
\end{align}
where the distortion $D$ satisfies the constraint \eqref{constraint-distortion-second-strategy} and $\boldsymbol{\alpha}_t$ should be chosen to minimize the effective noise $\dv z_t'$ in \eqref{effective-noise-maximum-ratio-combined-output-destination-second-strategy1}, i.e., such that 
\begin{align}
\boldsymbol{\alpha}_t^{\star}=\left(\dv G\dv G^T +\dv N_d\right)^{-1}\dv G \dv t,
\label{optimum-inflation-factors-decoding-at-destination-second-strategy1}
\end{align}
where $\dv G = [(\boldsymbol{\beta}_s \circ \mathrm{\mathbf{h}}_d)^T;\: (\boldsymbol{\beta}_s \circ\mathrm{\mathbf{h}}_r)^T]$ $\in \mathbb{R}^{2 \times 2}$ and $\dv N_d = [1/\text{snr},\:0;\:0,\: 1/\text{snr}+D/P]$ $\in \mathbb{R}^{2 \times 2}$. Similarly, in decoding the linear combination $\dv e_2 = k_a\dv v_a + k_b\dv v_b$, the probability of error at the destination $\text{Pr}(\mathrm{\mathbf{z}}_{k}' \notin \mc V_{\text{FINE}})$ goes to zero exponentially in $n$ if
\begin{align}
R &<\:\frac{1}{4}\:\log^{+}\left(\frac{\text{snr}}{\text{snr}||\boldsymbol{\beta}_s \circ \dv H^T\boldsymbol{\alpha}_k -\dv k||^2+(\boldsymbol{\alpha}_k \circ \boldsymbol{\alpha}_k)^T \dv n_d}\right)-\frac{1}{4}\log\left(2\pi e G(\Lambda)\right),
\label{second-computation-rate-destination-second-strategy}
\end{align}
where the distortion $D$ satisfies the constraint \eqref{constraint-distortion-second-strategy}, and $\boldsymbol{\alpha}_k$ should be chosen to minimize the effective noise $\dv z_k'$ in \eqref{effective-noise-maximum-ratio-combined-output-destination-second-strategy2}, i.e., such that 
\begin{align}
\boldsymbol{\alpha}_k^{\star}=\left(\dv G\dv G^T +\dv N_d\right)^{-1}\dv G \dv k.
\label{optimum-inflation-factors-decoding-at-destination-second-strategy2}
\end{align}

Let us define $R_1(\Lambda)$ as the RHS of \eqref{first-computation-rate-destination-second-strategy} and $R_2(\Lambda)$ as the RHS of \eqref{second-computation-rate-destination-second-strategy}. The above means that using the lattice-based coding scheme that we described, the destination can decode the transmitters' codewords correctly at the transmission symmetric-rate $R^{\text{CoD}}_{\text{sym}}(\Lambda) = \min\{R_1(\Lambda), R_2(\Lambda)\}$ provided that the condition \eqref{constraint-distortion-second-strategy} is satisfied. Furthermore, investigating the expression of $R_1(\Lambda)$, it can easily be seen that it decreases with increasing $D$. Also, observing that the RHS of \eqref{constraint-distortion-second-strategy} decreases if $\beta_r$ increases, the largest rate $R_1(\Lambda)$ is then obtained by taking the equality in the distortion constraint \eqref{constraint-distortion-second-strategy} with $\beta^2_r=P_r/P$. Finally, observing that $2\pi e G(\Lambda) \to 
 1$ when $n \to \infty$ \cite{CS88}, the desired symmetric-rate \eqref{achievable-sum-rate-compress-and-forward-at-relay-and-compute-at-destination} is obtained by taking the limit of $R^{\text{CoD}}_{\text{sym}}(\Lambda)$ as $n$ goes to infinity; and this completes the proof of Proposition~\ref{proposition-achievable-sum-rate-compress-and-forward-at-relay-and-compute-at-destination}. \qed

\begin{remark}\label{remark-local-vs-distributed-computation}
There are some high level similarities among the coding strategies of proposition~\ref{proposition-achievable-sum-rate-compute-and-forward-at-relay} and proposition~\ref{proposition-achievable-sum-rate-compress-and-forward-at-relay-and-compute-at-destination}. In particular, they both consist essentially in decoding two linearly independent equations. However, as we mentioned previously, the required two equations are obtained differently in the two cases. More specifically, while the two equations are computed in a distributed manner using the coding strategy of proposition~\ref{proposition-achievable-sum-rate-compute-and-forward-at-relay}, they are both computed locally at the destination in a joint manner using the coding strategy of proposition~\ref{proposition-achievable-sum-rate-compress-and-forward-at-relay-and-compute-at-destination}. A direct consequence of all the processing being performed locally at the destination with the latter coding scheme is that both computations of the required equations utilize \textit{all} the output available at the destination, i.e., the output received during the first transmission period as well as the output received during the second transmission period, in a joint manner. Recall that, by opposition, the coding strategy of proposition~\ref{proposition-achievable-sum-rate-compute-and-forward-at-relay} is such that the computation of one equation utilizes only the output received directly from the transmitters during the first transmission period, and the computation of the other equation is limited by the weaker output among the output at the relay during the first transmission period and the output at the destination during the second transmission period (since the equation decoded at the relay has to be recovered at the destination). The joint processing at the destination somehow gives some advantage to the coding strategy of proposition~\ref{proposition-achievable-sum-rate-compress-and-forward-at-relay-and-compute-at-destination}. The reader may refer to Section \ref{secV} where this aspect will be illustrated through some numerical examples and discussed further.
\end{remark}

\begin{remark}\label{remark-local-computation-vs-CF}
For the multiaccess relay network that we study, the coding strategy of Proposition~\ref{proposition-achievable-sum-rate-compress-and-forward-at-relay-and-compute-at-destination} can at best achieve the same performance as that allowed by regular compress-and-forward. This can be observed as follows. After conveying a quantized version of the relay's output to the destination, the decoding problem at the destination is equivalent to that over a regular two-user multiaccess channel with the output at the receiver given by $(\hat{\dv y}_r, \dv y_d)$. Optimal decoding of the messages can then be accomplished directly using joint decoding of the messages as in the CF-based approach of Section~\ref{secII_subsecC}. However, note that even though the coding strategy of Proposition~\ref{proposition-achievable-sum-rate-compress-and-forward-at-relay-and-compute-at-destination} can not achieve larger rates, it has some advantages over standard CF. For instance,  it is based on feasible structured codes instead of random codes which are infeasible in practice. Also, it utilizes linear receivers such as the decorrelator and minimum-mean-squared error receiver which are often used as low-complexity alternatives instead of maximum likelihood receiver which has high computational complexity. From this angle, note that this work also connects with \cite{SD11} in which the authors show that, for the standard Gaussian three-terminal relay channel, the rate achievable using standard CF can also be achieved alternately using lattice codes.
\end{remark}

%
%


\section{Symmetric Rates Optimization}\label{secIV}

Section~\ref{secIV_subsecA} is devoted to finding optimal powers and integer-coefficients that maximize the symmetric-rate of Proposition~\ref{proposition-achievable-sum-rate-compute-and-forward-at-relay}. Section~\ref{secIV_subsecB} deals with the optimization problem of Proposition~\ref{proposition-achievable-sum-rate-compress-and-forward-at-relay-and-compute-at-destination}.

\subsection{Compute-and-Forward at Relay}\label{secIV_subsecA}

\subsubsection{Problem Formulation}\label{secIV_subsecA_subsubsec1}

Consider the symmetric-rate $R^{\text{CoF}}_{\text{sym}}$ as given by \eqref{achievable-sum-rate-compute-and-forward-at-relay} in Proposition~\ref{proposition-achievable-sum-rate-compute-and-forward-at-relay}. The optimization problem can be stated as: 
\begin{eqnarray}
\text{(A)\:\: :} & \max\: \frac{1}{4} \min\left\{\log^{+}\left(\left(\|\mathrm{\mathbf{t}}\|^2-\frac{P((\boldsymbol{\beta}_s \circ \mathrm{\mathbf{h}}_d)^T \mathrm{\mathbf{t}})^2}{N+P\|\boldsymbol{\beta}_s \circ \mathrm{\mathbf{h}}_d\|^2}\right)^{-1}\right)\right.,\:\left.\log^{+}\left(\left(\|\mathrm{\mathbf{k}}\|^2-\frac{P((\boldsymbol{\beta}_s \circ \mathrm{\mathbf{h}}_r)^T \mathrm{\mathbf{k}})^2}{N+P\|\boldsymbol{\beta}_s \circ \mathrm{\mathbf{h}}_r\|^2}\right)^{-1}\right),\: \log\left(1+\frac{P|h_{rd}|^2\beta^2_r}{N}\right)\right\},
\label{statement-optimization-problem-first-strategy}
\end{eqnarray}
where the maximization is over $\boldsymbol{\beta}$ such that $0 \leq |\beta_a| \leq \sqrt{P_a/P}$, $0 \leq |\beta_b| \leq \sqrt{P_b/P}$, $0 \leq |\beta_r| \leq \sqrt{P_r/P}$ and over the integer coefficients $\dv k $ and $\dv t$ such that $|\text{det}(\dv k, \dv t)| \geq 1$.

The optimization problem (A) is non-linear and non-convex. Also, it is a MIQP optimization problem; and, so, it is not easy to solve it optimally. In what follows, we solve this optimization problem iteratively, by finding appropriate preprocessing vector $\mathrm{\boldsymbol{\beta}}$ and integer coefficients $\dv t$ and $\dv k$ alternately. We note that the allocation of the vector $\boldsymbol{\beta}$ determines the power that each of the transmitters and the relay should use for the transmission. For this reason, we will sometimes refer loosely to the process of selecting the vector $\boldsymbol{\beta}$ appropriately as the power allocation process. 

\noindent Let, with a slight abuse of notation, denote by $R^{\text{CoF}}_{\text{sym}}[\iota]$ the value of the symmetric-rate at some iteration $\iota \geq 0$. To compute $R^{\text{CoF}}_{\text{sym}}$ as given by  (\ref{statement-optimization-problem-first-strategy}) iteratively, we develop the following algorithm, to which we refer to as ``Algorithm A" in reference to the optimization problem (A).

\begin{algorithm}[h!] 
\renewcommand{\thealgorithm}{A}
{\fontsize{9}{9}\selectfont
\caption{\small{Iterative algorithm for computing $R^{\text{CoF}}_{\text{sym}}$ as given by (\ref{statement-optimization-problem-first-strategy})}}\label{al:1}
\begin{algorithmic}[1]
\State Initialization: set $\iota =1$ and $\mathrm{\boldsymbol{\beta}}=\mathrm{\boldsymbol{\beta}}^{(0)}$
\State Set $\mathrm{\boldsymbol{\beta}}=\mathrm{\boldsymbol{\beta}}^{(\iota-1)}$ in \eqref{statement-optimization-problem-first-strategy}, and solve the obtained problem using Algorithm A-1 given below. Denote by $\mathrm{\mathbf{k}}^{(\iota)}$ the found  $\mathrm{\mathbf{k}}$, and by $\mathrm{\mathbf{t}}^{(\iota)}$ the found $\mathrm{\mathbf{t}}$
\State Set $\mathrm{\mathbf{k}}=\mathrm{\mathbf{k}}^{(\iota)}$ and $\mathrm{\mathbf{t}}=\mathrm{\mathbf{t}}^{(\iota)}$ in \eqref{statement-optimization-problem-first-strategy}, and solve the obtained problem using Algorithm A-2 given below. Denote by $\boldsymbol{\beta}^{(\iota)}$ the found $\boldsymbol{\beta}$
\State Increment the iteration index as $\iota=\iota+1$, and go back to Step 2
\State Terminate if $\|\mathrm{\boldsymbol{\beta}}^{(\iota)}-\mathrm{\boldsymbol{\beta}}^{(\iota-1)}\|\leq \epsilon_1$, $|R^{\text{CoF}}_{\text{sym}}[\iota]-R^{\text{CoF}}_{\text{sym}}[\iota-1]|\leq \epsilon_2$
\end{algorithmic}
}
\end{algorithm}

As described in Algorithm A, we compute the appropriate preprocessing vector $\boldsymbol{\beta}$ and integer coefficients $\dv k$ and $\dv t$, alternately. More specifically, at iteration $\iota \geq 1$, the algorithm computes appropriate integer coefficients  $\mathrm{\mathbf{k}}^{(\iota)} \in \mathbb{Z}^2$ and $\mathrm{\mathbf{t}}^{(\iota)} \in \mathbb{Z}^2$ that correspond to a maximum of \eqref{statement-optimization-problem-first-strategy} computed with the choice of the preprocessing vector $\boldsymbol{\beta}$ set to its value obtained from the previous iteration, i.e., $\boldsymbol{\beta}=\boldsymbol{\beta}^{(\iota-1)}$ (for the initialization, set $\boldsymbol{\beta}^{(0)}$ to a default value). As we will show, this sub-problem is a MIQP problem with quadratic constraints; and we solve it iteratively using Algorithm A-1. Next, for the found integer coefficients, the algorithm computes adequate preprocessing vector $\boldsymbol{\beta}^{(\iota)}$ that corresponds to a maximum of \eqref{statement-optimization-problem-first-strategy} computed with the choice $\dv k=\mathrm{\mathbf{k}}^{(\iota)}$ and $\dv t=\mathrm{\mathbf{t}}^{(\iota)}$. As we will show, this sub-problem can be formulated as a complementary geometric programming problem. We solve it through a geometric programming and successive convex optimization approach (see Algorithm A-2 below).  The iterative process in Algorithm A terminates if the following two conditions hold: $\|\mathrm{\boldsymbol{\beta}}^{(\iota)}-\mathrm{\boldsymbol{\beta}}^{(\iota-1)}\|$ and $|R^{\text{CoF}}_{\text{sym}}[\iota]-R^{\text{CoF}}_{\text{sym}}[\iota-1]|$ are smaller than prescribed small strictly positive constants $\epsilon_1$ and $\epsilon_2$, respectively --- in this case, the optimized value of the symmetric-rate is $R^{\text{CoF}}_{\text{sym}}[\iota]$, and is attained using the preprocessing power vector $\boldsymbol{\beta}^{\star}=\mathrm{\boldsymbol{\beta}}^{(\iota)}$ and integer vectors $\dv k^{\star}= \mathrm{\mathbf{k}}^{(\iota)}$ and $\dv t^{\star}= \mathrm{\mathbf{t}}^{(\iota)}$.

In the following two sections, we study the aforementioned two sub-problems of problem (A), and describe the algorithms that we propose to solve them.

\vspace{0.3cm}

\subsubsection{Integer Coefficients Optimization}\label{secIV_subsecA_subsubsec2}

In this section, we focus on the problem of finding appropriate integer vectors $\dv k \in \mathbb{Z}^2$ and $\dv t \in \mathbb{Z}^2$ for a given choice of the preprocessing vector $\boldsymbol{\beta}$. Investigating the objective function in \eqref{statement-optimization-problem-first-strategy}, it can easily be seen that this problem can be equivalently stated as 
\begin{subequations}
\begin{eqnarray}
 \min_{\mathrm{\mathbf{k}},\: \mathrm{\mathbf{t}},\: \Delta_1}&&\Delta_1\\
\textrm{s. t.}&& \Delta_1 \geq \|\mathrm{\mathbf{t}}\|^2-\frac{P((\boldsymbol{\beta}_s \circ \mathrm{\mathbf{h}}_d)^T \mathrm{\mathbf{t}})^2}{N+P\|\boldsymbol{\beta}_s \circ \mathrm{\mathbf{h}}_d\|^2}\label{ICA1}\\
&& \Delta_1 \geq \|\mathrm{\mathbf{k}}\|^2-\frac{P((\boldsymbol{\beta}_s \circ \mathrm{\mathbf{h}}_r)^T \mathrm{\mathbf{k}})^2}{N+P\|\boldsymbol{\beta}_s \circ \mathrm{\mathbf{h}}_r\|^2}\label{ICA2}\\
&& \Delta_1 \geq \frac{N}{N+\beta^2_r|h_{rd}|^2P},\label{ICA3}\\
&& |\text{det}(\dv k, \dv t)| \geq 1\label{ICA4}\\
&& \mathrm{\mathbf{k}} \in \mathbb{Z}^2,\: \mathrm{\mathbf{t}} \in \mathbb{Z}^2,\:\: \Delta_1 \in \mathbb{R}.
\end{eqnarray}
\label{optimizing-integer-coefficients-for-given-beta}
\end{subequations}
Note that $\Delta_1$ is simultaneously an extra optimization variable and the objective function in \eqref{optimizing-integer-coefficients-for-given-beta}. Also, it is easy to see that the integer coefficients $\dv k$ and $\dv t$ that achieve the minimum value of $\Delta_1$ also achieve a maximum value of the objective function in \eqref{statement-optimization-problem-first-strategy}.  

In order to reformulate problem \eqref{optimizing-integer-coefficients-for-given-beta} in a manner that will be convenient for solving it, we introduce the following quantities. Let $\dv a_0=[0, 0, 0, 0, 1]^T$; $\dv a_1=\dv a_2=\dv a_3=[0, 0, 0, 0, -1]^T$ and $\dv a_4=[0, 0, 0, 0, 0]^T$. Also, let $\dv b=[t_a, t_b, k_a, k_b, \Delta_1]^T$; and the scalars $c_1=c_2=0$, $c_3=N/(N+\beta^2_r|h_{rd}|^2P)$ and $c_4=-1$. We also introduce the following five-by-five matrices $\dv F_1$, $\dv F_2$, $\dv F_3$ and $\dv F_4$, where
\begin{equation}
\dv F_1 =  \left[\begin{array}{cc}
2(\dv I_2 - \boldsymbol{\Omega}_1) & \dv 0 \\
\dv 0 & \dv 0
\end{array} \right], \qquad \dv F_2 = \left[\begin{array}{ccc}
\dv 0 & \dv 0 & \dv 0 \\
\dv 0 & 2(\dv I_2 - \boldsymbol{\Omega}_2)& \dv 0\\
 \dv 0 & \dv 0 & 0
\end{array} \right],
\end{equation}
with 
\begin{align}
\boldsymbol{\Omega}_1 &:= \frac{P}{N+P\|\dv h_d\|^2} (\boldsymbol{\beta}_s \circ \dv h_d)(\boldsymbol{\beta}_s \circ \dv h_d)^T,\nonumber\\
\boldsymbol{\Omega}_2 &:= \frac{P}{N+P\|\dv h_r\|^2} (\boldsymbol{\beta}_s \circ \dv h_r)(\boldsymbol{\beta}_s \circ \dv h_r)^T,
\end{align}
\begin{equation}
\dv F_3 = \dv 0, \:\:\textrm{and}\:\: \dv F_4 =  \left[\begin{array}{ccccc}
 \dv 0 & 0 & -2 & 0\\
 \dv 0 & 2 &  0 & 0\\
 \dv 0 & \dv 0 & \dv 0 & \dv 0\\
\end{array}  \right].
\end{equation}

\noindent The optimization problem \eqref{optimizing-integer-coefficients-for-given-beta} can now be reformulated equivalently as 
\begin{eqnarray}
\min_{\mathrm{\mathbf{b}}} && \dv a^T_0\dv b\nonumber \\
\textrm{s. t.}&& \frac{1}{2} \dv b^T\dv F_i\dv b+ \dv a^T_i \dv b \leq c_i\:\:\: i=1,\hdots,4 \nonumber\\
&& \dv k\in \mathbb{Z}^2,\: \dv t \in \mathbb{Z}^2,\:\: \Delta_1 \in \mathbb{R}
\label{equivalent-form-optimizing-integer-coefficients-for-given-beta}
\end{eqnarray}

\noindent The equivalent optimization problem \eqref{equivalent-form-optimizing-integer-coefficients-for-given-beta} is a MIQP problem with quadratic constraints \cite{F95}. If the involved matrices associated with the quadratic constraints (i.e., the matrices $\dv F_1$, $\dv F_2$ and $\dv F_4$ here) are all semi-definite, there are known approaches for solving MIQP optimization problems, such as cutting plane, decomposition, logic-based and branch and bound approaches \cite{F95}. In our case, it is easy to see that the matrices $\dv F_1$ and $\dv F_2$ are positive semi-definite. However, the matrix $\dv F_4$ is indefinite, irrespective to the values of $\dv k$ and $\dv t$. 

In order to transform the optimization problem \eqref{equivalent-form-optimizing-integer-coefficients-for-given-beta} into one that is MIQP-compatible (i.e., in which all the quadratic constraints are associated with semi-definite matrices), we replace the quadratic constraint \eqref{ICA4} with one that is linear, as follows. We introduce the following two real-valued vectors $\mathrm{\mathbf{\tilde{k}}}=[\tilde{k}_a, \tilde{k}_b]^T \in \mathbb{R}^2$ and $\mathrm{\mathbf{\tilde{t}}}=[\tilde{t}_a, \tilde{t}_b]^T \in \mathbb{R}^2$ defined such that they satisfy
\begin{align}
\mathrm{\mathbf{k}} = \boldsymbol{\kappa} \circ \exp(\mathrm{\mathbf{\tilde{k}}}), \quad \mathrm{\mathbf{t}} = \boldsymbol{\tau} \circ \exp(\mathrm{\mathbf{\tilde{t}}}),
\label{variables-changement-algorithm-A1}
\end{align}
where $\boldsymbol{\kappa}=[\kappa_a, \kappa_b]^T \in \mathbb{R}^2$ and $\boldsymbol{\tau}=[\tau_a, \tau_b]^T \in \mathbb{R}^2$ are constant vectors to be chosen appropriately. Thus, the constraint \eqref{ICA4} can now be rewritten equivalently as
\begin{align}
|\text{det}(\dv k, \dv t)| \geq 1 \quad &\text{iff} \qquad |\kappa_a\tau_b\text{exp}(\tilde{k}_{a}+\tilde{t}_{b})-\kappa_b\tau_a\text{exp}(\tilde{k}_{b}+\tilde{t}_{a})| \geq 1.
\label{equivalent-constraint-determinant-first-strategy}
\end{align}

\noindent Now, we linearize the constraint \eqref{equivalent-constraint-determinant-first-strategy} by selecting the constant vectors $\boldsymbol{\kappa}$ and $\boldsymbol{\tau}$ such that the first order Taylor series approximations $\exp(\mathrm{\mathbf{\tilde{k}}}) \approx \dv 1 + \mathrm{\mathbf{\tilde{k}}}$ and $\exp(\mathrm{\mathbf{\tilde{t}}}) \approx \dv 1 + \mathrm{\mathbf{\tilde{t}}}$ hold. Hence, the constraint \eqref{ICA4} can be rewritten as
\begin{align}
|\text{det}(\dv k, \dv t)| \geq 1 \quad &\text{iff} \qquad |\kappa_a\tau_b(1+\tilde{k}_a+\tilde{t}_b)-\kappa_b\tau_a(1+\tilde{k}_b+\tilde{t}_a)| \gtrsim 1.
\label{equivalent-form-constraint-determinant-first-strategy}
\end{align}

\noindent Note that the constraint \eqref{equivalent-form-constraint-determinant-first-strategy} is now linear. The optimization problem \eqref{equivalent-form-optimizing-integer-coefficients-for-given-beta}  has the same solution as the following problem which is MIQP-compatible,
\begin{subequations}
\begin{eqnarray}
 \min_{\mathrm{\mathbf{k}},\: \mathrm{\mathbf{t}},\: \Delta_1}&&\Delta_1\\
\textrm{s. t.}&& \|\mathrm{\mathbf{t}}\|^2-\frac{P((\boldsymbol{\beta}_s \circ \mathrm{\mathbf{h}}_d)^T \mathrm{\mathbf{t}})^2}{N+P\|\boldsymbol{\beta}_s \circ \mathrm{\mathbf{h}}_d\|^2} -\Delta_1 \leq 0\\ 
&& \|\mathrm{\mathbf{k}}\|^2-\frac{P((\boldsymbol{\beta}_s \circ \mathrm{\mathbf{h}}_r)^T \mathrm{\mathbf{k}})^2}{N+P\|\boldsymbol{\beta}_s \circ \mathrm{\mathbf{h}}_r\|^2} -\Delta_1 \leq 0\\
&& \frac{N}{N+\beta^2_r|h_{rd}|^2P} -\Delta_1 \leq 0\\
\label{final-form-optimizing-integer-coefficients-for-given-beta-determinant-constraint}
&& -|\kappa_a\tau_b(1+\tilde{k}_a+\tilde{t}_b)-\kappa_b\tau_a(1+\tilde{k}_b+\tilde{t}_a)| \lesssim -1\\
&&\frac{k_i}{\kappa_i} - 1 -\tilde{k}_i \leq 0, \quad -\frac{k_i}{\kappa_i} + 1 + \tilde{k}_i \leq 0, \quad i=a,b\\
&&\frac{t_i}{\tau_i} - 1 -\tilde{t}_i \leq 0, \quad -\frac{t_i}{\tau_i} + 1 + \tilde{t}_i \leq 0, \quad i=a,b\\
&&\dv k,\: \dv t \in \mathbb{Z}^2, \:\: \mathrm{\mathbf{\tilde{k}}}, \: \mathrm{\mathbf{\tilde{t}}},\: \boldsymbol{\kappa},\: \boldsymbol{\tau} \in \mathbb{R}^2, \Delta_1 \in \mathbb{R}.
\end{eqnarray}
\label{final-form-optimizing-integer-coefficients-for-given-beta}
\end{subequations}
The optimization problem \eqref{final-form-optimizing-integer-coefficients-for-given-beta} can be solved iteratively using Algorithm A-1 hereinafter.

\begin{algorithm}[h!]
\renewcommand{\thealgorithm}{A-1}
{\fontsize{9}{9}\selectfont
\caption{\small{Integer coefficients selection for $R^{\text{CoF}}_{\text{sym}}$ as given by (\ref{statement-optimization-problem-first-strategy})}}
\begin{algorithmic}[1]
\State Initialization: set $\iota_1 =1$ 
\State Use the branch-and-bound algorithm of \cite{LW66,W98} to solve for $\Delta_1^{(\iota_1)}$, $\mathrm{\mathbf{k}}^{(\iota_1)}$ and $\mathrm{\mathbf{t}}^{(\iota_1)}$ with the constraint \eqref{final-form-optimizing-integer-coefficients-for-given-beta-determinant-constraint} substituted with $-\kappa_a\tau_b(1+\tilde{k}_{a}+\tilde{t}_{b})+\kappa_b\tau_a(1+\tilde{k}_{b}+\tilde{t}_{a}) \leq -1$
\State Update the values of $\boldsymbol{\kappa}$ and $\boldsymbol{\tau}$ in a way to satisfy \eqref{variables-changement-algorithm-A1}; and increment the iteration index as $\iota_1=\iota_1+1$
\State Terminate if $\exp(\mathrm{\mathbf{\tilde{k}}}^{(\iota_1)}) \approx \dv 1 + \mathrm{\mathbf{\tilde{k}}}^{(\iota_1)}$ and $\exp(\mathrm{\mathbf{\tilde{t}}}^{(\iota_1)}) \approx \dv 1 + \mathrm{\mathbf{\tilde{t}}}^{(\iota_1)}$. Denote the found solution as $\Delta_1^{\text{min},1}$
\State Redo steps 1 to 4 with in Step 2 the constraint \eqref{final-form-optimizing-integer-coefficients-for-given-beta-determinant-constraint} substituted with $\kappa_a\tau_b(1+\tilde{k}_{a}+\tilde{t}_{b})-\kappa_b\tau_a(1+\tilde{k}_{b}+\tilde{t}_{a}) \leq -1$. In Step 4, denote the found solution as $\Delta_1^{\text{min},2}$
\State Select the integer coefficients corresponding to the minimum among $\Delta_1^{\text{min},1}$ and $\Delta_1^{\text{min},2}$
\end{algorithmic}
}
\end{algorithm}

\subsubsection{Power Allocation Policy}\label{secIV_subsecA_subsubsec3}

Let us now focus on the problem of finding an appropriate preprocessing vector $\boldsymbol{\beta}$ for given integer vectors $\dv k \in \mathbb{Z}^2$ and $\dv t \in \mathbb{Z}^2$. Again, investigating the objective function in \eqref{statement-optimization-problem-first-strategy}, it can easily be seen that this problem can be equivalently stated as
\begin{subequations}
\begin{eqnarray}
\min_{\mathrm{\boldsymbol{\beta}},\:\Delta_2}&& \Delta_2 \\
\textrm{s. t.}&& \Delta_2 \geq \|\mathrm{\mathbf{t}}\|^2-\frac{P((\boldsymbol{\beta}_s \circ \mathrm{\mathbf{h}}_d)^T \mathrm{\mathbf{t}})^2}{N+P\|\boldsymbol{\beta}_s \circ \mathrm{\mathbf{h}}_d\|^2} \label{power_allocation_1}\\
&& \Delta_2 \geq \|\mathrm{\mathbf{k}}\|^2-\frac{P((\boldsymbol{\beta}_s \circ \mathrm{\mathbf{h}}_r)^T \mathrm{\mathbf{k}})^2}{N+P\|\boldsymbol{\beta}_s \circ \mathrm{\mathbf{h}}_r\|^2} \label{power_allocation_2}\\
&& \Delta_2 \geq \frac{N}{N+P|h_{rd}|^2\beta^2_r}, \label{power_allocation_3}\\
&& -\sqrt{\frac{P_i}{P}} \leq \beta_i \leq \sqrt{\frac{P_i}{P}},\:\: i = a,b,r\label{power_allocation_4}\\
&& \boldsymbol{\beta}_s \in \mathbb{R}^2,\: \boldsymbol{\beta} \in \mathbb{R}^3,\: \Delta_2 \in \mathbb{R}.
\end{eqnarray}
\label{optimizing-betas-for-given-integer-coefficients}
\end{subequations}
Here, similarly to the previous section, $\Delta_2$ is simultaneously an extra optimization variable and the objective function in \eqref{optimizing-betas-for-given-integer-coefficients}. Also, it is easy to see that the value of $\boldsymbol{\beta}$ that achieves the minimum value of $\Delta_2$ also achieves a maximum value of the objective function in \eqref{statement-optimization-problem-first-strategy}. 

\noindent The optimization problem in \eqref{optimizing-betas-for-given-integer-coefficients} is non-linear and non-convex. We use geometric programming \cite{CTPOJ07} for solving it. Geometric programming is a special form of convex optimization for which efficient algorithms have been developed and are known in the related literature \cite{C05}. There are two forms of GP: the standard form and the convex form. In its standard form, a GP optimization problem is generally written as \cite{C05}
\begin{subequations}
\begin{eqnarray}\label{GP}
\textrm{minimize}&& f_0(\mathrm{\boldsymbol{\beta}},\Delta_2)\\
\textrm{subject to}&& f_j(\mathrm{\boldsymbol{\beta}},\Delta_2)\leq 1, \:\:\:\:\: j =1,\hdots,J,\\
&& g_l(\mathrm{\boldsymbol{\beta}},\Delta_2)= 1, \:\:\:\:\: l =1,\hdots, L,
\end{eqnarray}
\label{standard-form-of-geometric-programming}
\end{subequations}
where the functions $f_0$ and $f_j$, $j=1,\hdots,J$, are posynomials and the functions $g_l$, $l=1,\hdots,L$, are monomials in $\mathrm{\boldsymbol{\beta}}$ and $\Delta_2$. In its standard form, \eqref{standard-form-of-geometric-programming} is not a convex optimization problem. However, when possible, a careful application of an appropriate logarithmic transformation of the involved variables and constants generally turns the problem \eqref{standard-form-of-geometric-programming} into one that is equivalent and convex. That is, \eqref{standard-form-of-geometric-programming} is a GP nonlinear, nonconvex optimization problem that can be transformed into a nonlinear, convex optimization problem.

In our case, in the problem \eqref{optimizing-betas-for-given-integer-coefficients}, the constraints \eqref{power_allocation_1} and \eqref{power_allocation_2} contain functions that are non posynomial. Also, the variables in \eqref{optimizing-betas-for-given-integer-coefficients} are not all positive, thus preventing a direct application of logarithmic transformation. In what follows, we first transform the problem \eqref{optimizing-betas-for-given-integer-coefficients} into one equivalent in which the constraints involve functions that are all posynomial and the variables are all positive; and then we develop an algorithm for solving the equivalent problem.

Let $\dv c=[c_a, c_b, c_r]^T \in \mathbb{R}^3$ and $\boldsymbol{\delta}=[\delta_a, \delta_b, \delta_r]^T \in \mathbb{R}^3$, such that  $c_i > \sqrt{P_i/P}$ and $\delta_i=\beta_i+c_i$ for $i=a, b, r$. Note that the elements of $\boldsymbol{\delta}$ are all strictly positive. Also, for convenience, we define the following functions, for $\dv z=[z_a, z_b] \in \mathbb{Z}^2$,
\begin{align}
\psi_1^i(\mathrm{\boldsymbol{\delta}}, \Delta_2, \dv z) &= 2\Delta_2P\left(|h_{ai}|^2\delta_ac_a+|h_{bi}|^2\delta_bc_b\right)+P\left(z_a^2+z_b^2\right)\left(|h_{ai}|^2(\delta_a^2+c_a^2)+|h_{bi}|^2(\delta_b^2+c_b^2)\right)\nonumber\\
&\quad +2P\left(|h_{ai}|^2z_a^2\delta_ac_a+|h_{bi}|^2z_b^2\delta_bc_b+h_{ai}h_{bi}z_az_b(\delta_ac_b+\delta_bc_a)\right)+N\left(z_a^2+z_b^2\right)\nonumber\\
\psi_2^i(\mathrm{\boldsymbol{\delta}}, \Delta_2, \dv z) &= \Delta_2\left(N+P|h_{ai}|^2(\delta_a^2+c_a^2)+P|h_{bi}|^2(\delta_b^2+c_b^2)\right)+2P\left(z_a^2+z_b^2\right)\left(|h_{ai}|^2\delta_ac_a+|h_{bi}|^2\delta_bc_b\right)\nonumber\\
&\quad +P\left(|h_{ai}|^2z_a^2(\delta_a^2+c_a^2)+|h_{bi}|^2z_b^2(\delta_b^2+c_b^2)+2h_{ai}h_{bi}z_az_b(\delta_a\delta_b+c_ac_b)\right).
\end{align}

Let us now define the following functions, $f_1(\mathrm{\boldsymbol{\delta}}, \Delta_2) = \psi_1^d(\mathrm{\boldsymbol{\delta}},\Delta_2, \dv t)$, $\:f_2(\mathrm{\boldsymbol{\delta}}, \Delta_2)  = \psi_1^r(\mathrm{\boldsymbol{\delta}},\Delta_2,  \dv k)$, $\:g_1(\mathrm{\boldsymbol{\delta}}, \Delta_2) = \psi_2^d(\mathrm{\boldsymbol{\delta}},\Delta_2, \dv t)$, $\:g_2(\mathrm{\boldsymbol{\delta}}, \Delta_2) = \psi_2^r(\mathrm{\boldsymbol{\delta}},\Delta_2, \dv k)$, $\:f_3(\mathrm{\boldsymbol{\delta}},\Delta_2) = N+2\Delta_2P|h_{rd}|^2\delta_{r}c_r$, and $\:g_3(\mathrm{\boldsymbol{\delta}}, \Delta_2) = \Delta_2\left(N +P|h_{rd}|^2(\delta_{r}^2+c_r^2)\right)$.

It is now easy to see that the optimization problem \eqref{optimizing-betas-for-given-integer-coefficients} can be stated in the following form.
\begin{subequations}
\begin{eqnarray}
\min_{\mathrm{\boldsymbol{\delta}},\: \Delta_2}&&\Delta_2 \\
\label{constraints-equivalent-form-optimizing-betas-for-given-integer-coefficients}
\textrm{s. t.}&&\frac{f_1(\mathrm{\boldsymbol{\delta}},\Delta_2)}{g_1(\mathrm{\boldsymbol{\delta}},\Delta_2)}\leq 1,\quad  \frac{f_2(\mathrm{\boldsymbol{\delta}},\Delta_2)}{g_2(\mathrm{\boldsymbol{\delta}},\Delta_2)}\leq 1, \quad \frac{f_3(\mathrm{\boldsymbol{\delta}},\Delta_2)}{g_3(\mathrm{\boldsymbol{\delta}},\Delta_2)}\leq 1 \\
&& -\sqrt{\frac{P_i}{P}}+ c_i \leq \delta_i \leq \sqrt{\frac{P_i}{P}}+ c_i \:\:,\: i =a, b, r\\
&& \boldsymbol{\delta} \in \mathbb{R}^3,\:\: \dv c \in \mathbb{R}^3, \:\: \Delta_2 \in \mathbb{R}.
\end{eqnarray}
\label{equivalent-form-optimizing-betas-for-given-integer-coefficients}
\end{subequations}

The constraints \eqref{constraints-equivalent-form-optimizing-betas-for-given-integer-coefficients} involve functions that consist of ratios of posynomials, i.e., are not posynomial --- recall that a ratio of posynomials is in general non posynomial. Minimizing or upper bounding a ratio of posynomials belongs to a class of non-convex problems known as complementary  GP \cite{C05}. One can transform a complementary GP problem into a GP problem using series of approximations. In order to get posynomials, we approximate the functions $g_1(\mathrm{\boldsymbol{\delta}},\Delta_2)$, $g_2(\mathrm{\boldsymbol{\delta}},\Delta_2)$ and $g_3(\mathrm{\boldsymbol{\delta}},\Delta_2)$ with monomials, by using the following lemma \cite{CTPOJ07}.

\vspace{0.2cm}

\begin{lemma}\label{lemma1}
Let $g(\mathrm{\boldsymbol{\delta}},\Delta_2)= \sum_j{u_j(\mathrm{\boldsymbol{\delta}},\Delta_2)}$ be a posynomial. Then
\begin{eqnarray}
g(\mathrm{\boldsymbol{\delta}},\Delta_2) \geq \tilde{g}(\mathrm{\boldsymbol{\delta}},\Delta_2) = \prod_j{\left(\frac{u_j(\mathrm{\boldsymbol{\delta}},\Delta_2)}{\gamma_j}\right)^{\gamma_j}}.
\label{convex-approximations}
\end{eqnarray}
Here, $\gamma_j = u_j(\mathrm{\boldsymbol{\delta}}^{(0)},\Delta_2^{(0)})/g(\mathrm{\boldsymbol{\delta}}^{(0)},\Delta_2^{(0)})$, $\forall j$, for any fixed positive $\mathrm{\boldsymbol{\delta}}^{(0)}$ and $\Delta_2^{(0)}$ then $\tilde{g}(\mathrm{\boldsymbol{\delta}}^{(0)},\Delta_2^{(0)})=g(\mathrm{\mathbf{\delta}}^{(0)},\Delta_2^{(0)})$, and $\tilde{g}(\mathrm{\boldsymbol{\delta}}^{(0)},\Delta_2^{(0)})$ is the best local monomial approximation to $g(\mathrm{\boldsymbol{\delta}}^{(0)},\Delta_2^{(0)})$ near $\mathrm{\boldsymbol{\delta}}^{(0)}$ and $\Delta_2^{(0)}$.
\end{lemma}

\vspace{0.2cm}

Let $\tilde{g}_1(\mathrm{\boldsymbol{\delta}},\Delta_2)$, $\tilde{g}_2(\mathrm{\boldsymbol{\delta}},\Delta_2)$ and $\tilde{g}_3(\mathrm{\boldsymbol{\delta}},\Delta_2)$ be the monomial approximations of the functions $g_1(\mathrm{\boldsymbol{\delta}},\Delta_2)$, $g_2(\mathrm{\boldsymbol{\delta}},\Delta_2)$ and $g_3(\mathrm{\boldsymbol{\delta}},\Delta_2)$ obtained using Lemma~\ref{lemma1}. Using these monomial approximations, the ratios of posynomials involved in the constraint \eqref{constraints-equivalent-form-optimizing-betas-for-given-integer-coefficients} can be upper bounded by posynomials. The optimal solution of the problem obtained using the convex approximations is also optimal for the  original problem \eqref{optimizing-betas-for-given-integer-coefficients}, i.e., satisfies the Karush-Kuhn-Tucker (KKT) conditions of the original problem \eqref{optimizing-betas-for-given-integer-coefficients}, if the applied approximations satisfy the following three properties \cite{
 MW78, CTPOJ07}:

\begin{enumerate} 
\item[1)] $g_j(\mathrm{\boldsymbol{\delta}},\Delta_2)\leq \tilde{g}_j(\mathrm{\boldsymbol{\delta}},\Delta_2)$ for all $\mathrm{\boldsymbol{\delta}}$ and $\Delta_2$ where $\tilde{g}_j(\mathrm{\boldsymbol{\delta}},\Delta_2)$ is the approximation of ${g}_j(\mathrm{\boldsymbol{\delta}},\Delta_2)$.
\item[2)] $g_j(\mathrm{\boldsymbol{\delta}}^{(0)},\Delta_2^{(0)})= \tilde{g}_j(\mathrm{\boldsymbol{\delta}}^{(0)},\Delta_2^{(0)})$ where $\mathrm{\boldsymbol{\delta}}^{(0)}$ and $\Delta_2^{(0)}$ are the optimal solution of the approximated problem in the previous iteration.
\item[3)] $\triangledown g_j(\mathrm{\boldsymbol{\delta}}^{(0)},\Delta_2^{(0)}) = \bigtriangledown \tilde{g}_j(\mathrm{\boldsymbol{\delta}}^{(0)},\Delta_2^{(0)})$, where $\triangledown g_j(\cdot)$ stands for the gradient of function $g_j(\cdot)$.
\end{enumerate}

Summarizing, applying the aforementioned transformations, we transformed the original optimization problem \eqref{optimizing-betas-for-given-integer-coefficients} first into a complementary GP problem \eqref{equivalent-form-optimizing-betas-for-given-integer-coefficients} and then into a GP problem by applying the convex approximations \eqref{convex-approximations}. Finally, the obtained GP problem can be solved easily using, e.g., an interior point approach. More specifically, the problem of finding the appropriate preprocessing vector $\mathrm{\boldsymbol{\delta}}$ for given integer vectors $\dv k$ and $\dv t$ can be solved using Algorithm A-2 hereinafter. 

\begin{algorithm}[h!]
\renewcommand{\thealgorithm}{A-2}
{\fontsize{9}{9}\selectfont
\caption{\small{Power allocation policy for $R^{\text{CoF}}_{\text{sym}}$ as given by (\ref{statement-optimization-problem-first-strategy})}}
\begin{algorithmic}[1]
\State Set $\mathrm{\boldsymbol{\delta}}^{(0)}$ to some initial value. Compute $\Delta_2^{(0)}$ using $\mathrm{\boldsymbol{\delta}}^{(0)}$ and set $\iota_2 =1$
\State Approximate $g(\mathrm{\boldsymbol{\delta}}^{(\iota_2)},\Delta_2^{(\iota_2)})$ with $\tilde{g}(\mathrm{\boldsymbol{\delta}}^{(\iota_2)},\Delta_2^{(\iota_2)})$ around $\mathrm{\boldsymbol{\delta}}^{(\iota_2-1)}$ and $\Delta_2^{(\iota_2-1)}$ using \eqref{convex-approximations}
\State Solve the resulting approximated GP problem using an interior point approach. Denote the found solutions as $\mathrm{\boldsymbol{\delta}}^{(\iota_2)}$ and $\Delta_2^{(\iota_2)}$
\State Increment the iteration index as $\iota_2=\iota_2+1$ and go back to Step 2 using $\mathrm{\boldsymbol{\delta}}$ and $\Delta_2$ of step 3
\State Terminate if $\|\mathrm{\boldsymbol{\delta}}^{(\iota_2)}-\mathrm{\boldsymbol{\delta}}^{(\iota_2-1)}\|\leq \epsilon_1$
\end{algorithmic}
}
\end{algorithm}

\subsection{Compress-and-Forward at Relay and Compute at Destination}\label{secIV_subsecB}

The algorithms that we develop in this section to solve the optimization problem involved in the maximization of the symmetric-rate given in Proposition~\ref{proposition-achievable-sum-rate-compress-and-forward-at-relay-and-compute-at-destination} are essentially similar to those that we developed in the previous section. For brevity, we omit the details in this section.

\subsubsection{Problem Formulation}\label{secIV_subsecB_subsubsec1}

Recall the expression of $R^{\text{CoD}}_{\text{sym}}$ as given by \eqref{achievable-sum-rate-compress-and-forward-at-relay-and-compute-at-destination} in Proposition~\ref{proposition-achievable-sum-rate-compress-and-forward-at-relay-and-compute-at-destination}. The optimization problem can be stated as:

\begin{align}
\text{(B)\:\: :\:\:} \max \: \frac{1}{4}\min\Bigg\{&\log^{+}\left(\frac{\text{snr}}{\text{snr}||\boldsymbol{\beta}_s \circ \dv H^T\boldsymbol{\alpha}_t -\dv t||^2+(\boldsymbol{\alpha}_t \circ \boldsymbol{\alpha}_t)^T \dv n_d}\right),\nonumber\\
& \log^{+}\left(\frac{\text{snr}}{\text{snr}||\boldsymbol{\beta}_s \circ \dv H^T\boldsymbol{\alpha}_k -\dv k||^2+(\boldsymbol{\alpha}_k \circ \boldsymbol{\alpha}_k)^T \dv n_d}\right)\Bigg\},
\label{optimization-problem-compress-and-forward-at-relay-and-compute-at-destination}
\end{align}
where the distortion $D$ is given by
\begin{equation}
D = \frac{N^2\left(1+\text{snr}\|\boldsymbol{\beta_s} \circ \dv h_r\|^2\right)}{|h_{rd}|^2P_r}-\frac{N^2\big(\text{snr}(\boldsymbol{\beta_s}\circ\dv h_r)^T(\boldsymbol{\beta_s}\circ\dv h_d)\big)^2}{|h_{rd}|^2P_r\left(1+\text{snr}\|\boldsymbol{\beta_s} \circ \dv h_d\|^2\right)}, 
\end{equation}
and the maximization is over $\boldsymbol{\alpha}_t$, $\boldsymbol{\alpha}_k$, $\boldsymbol{\beta}_s$ such that $0 \leq |\beta_a| \leq \sqrt{P_a/P}$, $0 \leq |\beta_b| \leq \sqrt{P_b/P}$, and over the integer coefficients $\dv k$ and $\dv t$ such that $|\text{det}(\dv k, \dv t)| \geq 1$.

\noindent In order to compute $R^{\text{CoD}}_{\text{sym}}$ as given by (\ref{optimization-problem-compress-and-forward-at-relay-and-compute-at-destination}), we develop the following iterative algorithm which optimizes the integer coefficients and the powers alternately, and to which we refer to as ``Algorithm B" in reference to the optimization problem (B).

\begin{algorithm}[h!]
\renewcommand{\thealgorithm}{B}
{\fontsize{9}{9}\selectfont
\caption{\small{Iterative algorithm for computing $R^{\text{CoD}}_{\text{sym}}$ as given by (\ref{optimization-problem-compress-and-forward-at-relay-and-compute-at-destination})}} 
\begin{algorithmic}[1]
\State Choose an initial feasible vector $\mathrm{\boldsymbol{\beta}_s}^{(0)}$ and set $\iota =1$
\State Solve \eqref{optimization-problem-compress-and-forward-at-relay-and-compute-at-destination} with $\mathrm{\boldsymbol{\beta}_s}=\mathrm{\boldsymbol{\beta}_s}^{(\iota-1)}$ for the optimal $\mathrm{\mathbf{k}}$ and $\mathrm{\mathbf{t}}$ using Algorithm B-1 and assign it to $\mathrm{\mathbf{k}}^{(\iota)}$ and $\mathrm{\mathbf{t}}^{(\iota)}$
\State Solve \eqref{optimization-problem-compress-and-forward-at-relay-and-compute-at-destination} with $\mathrm{\mathbf{k}}=\mathrm{\mathbf{k}}^{(\iota)}$ and $\mathrm{\mathbf{t}}=\mathrm{\mathbf{t}}^{(\iota)}$ for the optimal $\mathrm{\boldsymbol{\beta}_s}$ using Algorithm B-2 and assign it to $\mathrm{\boldsymbol{\beta}_s}^{(\iota)}$
\State Increment the iteration index as $\iota=\iota+1$ and go back to Step 2
\State Terminate if $\|\mathrm{\boldsymbol{\beta}_s}^{(\iota)}-\mathrm{\boldsymbol{\beta}_s}^{(\iota-1)}\|\leq \epsilon_1$, $|R^{\text{CoD}}_{\text{sym}}[\iota]-R^{\text{CoD}}_{\text{sym}}[\iota-1]| \leq \epsilon_2$ 
\end{algorithmic}
}
\end{algorithm}

\subsubsection{Integer Coefficients Optimization}\label{secIV_subsecB_subsubsec2}

Proceeding similarly as above, the problem of optimizing the integer vectors $\dv k$ and $\dv t$ for a fixed choice of the preprocessing vector $\mathrm{\boldsymbol{\beta}_s}$ can be written as
\begin{subequations}
\begin{eqnarray}
\min_{\mathrm{\mathbf{k}},\: \mathrm{\mathbf{t}},\: \Theta_1}&&\Theta_1\\
\textrm{s. t.}&& \Theta_1 \geq \dv t^T\boldsymbol{\Omega}\dv t\\
&&\Theta_1 \geq \dv k^T\boldsymbol{\Omega}\dv k\\
&& \text{det}(\dv k, \dv t) = |k_at_b-k_bt_a|\geq 1\label{ICB1}\\
&& \mathrm{\mathbf{k}},\: \mathrm{\mathbf{t}} \in \mathbb{Z}^2,\:\: \Theta_1 \in \mathbb{R},
\end{eqnarray}\label{optimizing-integer-coefficients-for-given-beta-B1}
\end{subequations}
where $\boldsymbol{\Omega} = (\dv G^T(\dv G\dv G^T+\dv N_d)^{-1}\dv G-\dv I_2)^T(\dv G^T(\dv G\dv G^T+\dv N_d)^{-1}\dv G-\dv I_2)+((\dv G\dv G^T+\dv N_d)^{-1}\dv G)^T \dv N_d((\dv G\dv G^T+\dv N_d)^{-1}\dv G)$.
We reformulate the problem \eqref{optimizing-integer-coefficients-for-given-beta-B1} into a MIQP problem with quadratic constraints \cite{F95} as before. We introduce the following quantities. Let $\dv a_0=[0, 0, 0, 0, 1]^T$; $\dv a_1=\dv a_2=[0, 0, 0, 0, -1]^T$ and $\dv a_3=[0, 0, 0, 0, 0]^T$. Also, let $\dv b=[t_a, t_b, k_a, k_b, \Theta_1]^T$; and the scalars $c_1=c_2=0$, and $c_3=-1$. We also introduce the following five-by-five matrices $\dv F_1$, $\dv F_2$, and $\dv F_3$, where
\begin{equation}
\dv F_1 =  \left[\begin{array}{cc}
2\boldsymbol{\Omega} & \dv 0 \\
\dv 0 & \dv 0
\end{array} \right], \qquad \dv F_2 = \left[\begin{array}{ccc}
\dv 0 & \dv 0 & \dv 0 \\
\dv 0 & 2\boldsymbol{\Omega}& \dv 0\\
 \dv 0 & \dv 0 &  0
\end{array} \right], \:\:\text{and}\:\:\dv F_3 =  \left[\begin{array}{ccccc}
 \dv 0 & 0 & -2 &  0\\
 \dv 0 & 2 &  0 &  0\\
 \dv 0 & \dv 0 &  \dv 0 & \dv 0\\
\end{array}  \right].
\end{equation}

\noindent The optimization problem \eqref{optimizing-integer-coefficients-for-given-beta-B1} can now be reformulated equivalently as 
\begin{eqnarray}
\min_{\mathrm{\mathbf{b}}} && \dv a^T_0\dv b\nonumber \\
\textrm{s. t.}&& \frac{1}{2} \dv b^T\dv F_i\dv b+ \dv a^T_i \dv b \leq c_i\:\:\: i=1,2,3 \nonumber\\
&& \dv k\in \mathbb{Z}^2,\: \dv t \in \mathbb{Z}^2,\:\: \Theta_1 \in \mathbb{R}
\label{equivalent-form-optimizing-integer-coefficients-for-given-beta-B1}
\end{eqnarray}

\noindent It is easy to see that the matrices $\dv F_1$ and $\dv F_2$ are positive semi-definite. However, the matrix $\dv F_3$ is indefinite, irrespective to the values of $\dv k$ and $\dv t$. 
In order to transform the optimization problem \eqref{equivalent-form-optimizing-integer-coefficients-for-given-beta-B1} into one that is MIQP-compatible (i.e., in which all the quadratic constraints are associated with semi-definite matrices), we replace the quadratic constraint \eqref{ICB1} with one that is linear, as performed in the previous section. 

Finally, the MIQP optimization problem is given by,
\begin{subequations}
\begin{eqnarray}\label{Int_Alloc_LCF}
\min_{\mathrm{\mathbf{k}},\: \mathrm{\mathbf{t}},\: \Theta_1}&&\Theta_1\\
\textrm{s. t.}&& \Theta_1 \geq \dv t^T\boldsymbol{\Omega}\dv t\\
&&\Theta_1 \geq \dv k^T\boldsymbol{\Omega}\dv k\\
\label{final-form-optimizing-integer-coefficients-for-given-beta-determinant-constraint-B1}
&& -|\kappa_a\tau_b(1+\tilde{k}_a+\tilde{t}_b)-\kappa_b\tau_a(1+\tilde{k}_b+\tilde{t}_a)| \lesssim -1\\
&&\frac{k_i}{\kappa_i} - 1 -\tilde{k}_i \leq 0, \quad -\frac{k_i}{\kappa_i} + 1 + \tilde{k}_i \leq 0, \quad i=a,b\\
&&\frac{t_i}{\tau_i} - 1 -\tilde{t}_i \leq 0, \quad -\frac{t_i}{\tau_i} + 1 + \tilde{t}_i \leq 0, \quad i=a,b\\
&&\:\:\dv k,\: \dv t \in \mathbb{Z}^2, \:\: \mathrm{\mathbf{\tilde{k}}}, \: \mathrm{\mathbf{\tilde{t}}} ,\: \boldsymbol{\kappa} , \: \boldsymbol{\tau} \in \mathbb{R}^2, \Theta_1 \in \mathbb{R}.
\end{eqnarray}
\label{final-form-optimizing-integer-coefficients-for-given-beta-B1}
\end{subequations}
The optimization problem \eqref{final-form-optimizing-integer-coefficients-for-given-beta-B1} can be solved iteratively using Algorithm B-1 hereinafter.

\begin{algorithm}[h!]
\renewcommand{\thealgorithm}{B-1}
\caption{\small{Integer coefficients selection for $R^{\text{CoD}}_{\text{sym}}$ as given by (\ref{optimization-problem-compress-and-forward-at-relay-and-compute-at-destination})}}
{\fontsize{9}{9}\selectfont
\begin{algorithmic}[1]
\State Initialization: set $\iota_1 =1$
\State Use the branch-and-bound algorithm of \cite{LW66,W98} to solve for $\Theta_1^{(\iota_1)}$, $\mathrm{\mathbf{k}}^{(\iota_1)}$ and $\mathrm{\mathbf{t}}^{(\iota_1)}$ with the constraint \eqref{final-form-optimizing-integer-coefficients-for-given-beta-determinant-constraint-B1} substituted with $-\kappa_a\tau_b(1+\tilde{k}_{a}+\tilde{t}_{b})+\kappa_b\tau_a(1+\tilde{k}_{b}+\tilde{t}_{a}) \leq -1$
\State Update the values of $\boldsymbol{\kappa}$ and $\boldsymbol{\tau}$ in a way to satisfy \eqref{variables-changement-algorithm-A1}; and increment the iteration index as $\iota_1=\iota_1+1$
\State Terminate if $\exp(\mathrm{\mathbf{\tilde{k}}}^{(\iota_1)}) \approx \dv 1 + \mathrm{\mathbf{\tilde{k}}}^{(\iota_1)}$ and $\exp(\mathrm{\mathbf{\tilde{t}}}^{(\iota_1)}) \approx \dv 1 + \mathrm{\mathbf{\tilde{t}}}^{(\iota_1)}$. Denote the found solution as $\Theta_1^{\text{min},2}$
\State Redo steps 1 to 4 with in Step 2 the constraint \eqref{final-form-optimizing-integer-coefficients-for-given-beta-determinant-constraint-B1} substituted with $\kappa_a\tau_b(1+\tilde{k}_{a}+\tilde{t}_{b})-\kappa_b\tau_a(1+\tilde{k}_{b}+\tilde{t}_{a}) \leq -1$. In Step 4, denote the found solution as $\Theta_1^{\text{min},2}$
\State Select the integer coefficients corresponding to the minimum among $\Theta_1^{\text{min},1}$ and $\Theta_1^{\text{min},2}$
\end{algorithmic}
}
\end{algorithm}

\subsubsection{Power Allocation Policy}\label{secIV_subsecB_subsubsec3}

The problem of optimizing the power value $\mathrm{\boldsymbol{\beta}_s}$ for a fixed integer coefficients $\mathrm{\mathbf{k}}$, and $\mathrm{\mathbf{t}}$, can be written as,
\begin{subequations}
\begin{eqnarray}
\min_{\mathrm{\boldsymbol{\beta}_s},\: \boldsymbol{\alpha}_t,\: \boldsymbol{\alpha}_k,\: \Theta_2}&&\Theta_2\\
\textrm{s. t.}&& \Theta_2 \geq \frac{\text{snr}||\boldsymbol{\beta}_s \circ \dv H^T\boldsymbol{\alpha}_t -\dv t||^2+(\boldsymbol{\alpha}_t \circ \boldsymbol{\alpha}_t)^T \dv n_d}{\text{snr}}\:,\label{PACF1}\\
&&\Theta_2 \geq \frac{\text{snr}||\boldsymbol{\beta}_s \circ \dv H^T\boldsymbol{\alpha}_k -\dv k||^2+(\boldsymbol{\alpha}_k \circ \boldsymbol{\alpha}_k)^T \dv n_d}{\text{snr}}\:,\label{PACF2}\\
&& D \geq \frac{N^2\left(1+\text{snr}\|\boldsymbol{\beta_s} \circ \dv h_r\|^2\right)}{|h_{rd}|^2P_r}-\frac{N^2\big(\text{snr}(\boldsymbol{\beta_s}\circ\dv h_r)^T(\boldsymbol{\beta_s}\circ\dv h_d)\big)^2}{|h_{rd}|^2P_r\left(1+\text{snr}\|\boldsymbol{\beta_s} \circ \dv h_d\|^2\right)} \label{PACF3}\\
&& -\sqrt{\frac{P_i}{P}} \leq \beta_i \leq \sqrt{\frac{P_i}{P}},\:\: i = a,b\\
&& \boldsymbol{\beta}_s \in \mathbb{R}^2,\: \Theta_2 \in \mathbb{R}.
\end{eqnarray}
\label{optimizing-betas-for-given-integer-coefficients-B2}
\end{subequations}
As before, let $\dv c=[c_a, c_b]^T \in \mathbb{R}^2$ and $\boldsymbol{\delta}_s=[\delta_a, \delta_b]^T \in \mathbb{R}^2$, such that  $c_i > \sqrt{P_i/P}$ and $\delta_i=\beta_i+c_i$ for $i=a, b$. Note that the elements of $\boldsymbol{\delta}_s$ are all strictly positive. We can reformulate the optimization problem as,
\begin{subequations}
\begin{eqnarray}
\min_{\mathrm{\boldsymbol{\delta_s}},\: \boldsymbol{\alpha}_t,\: \boldsymbol{\alpha}_k,\: \Theta_2}&&\Theta_2 \\
\label{constraints-equivalent-form-optimizing-betas-for-given-integer-coefficients-B2}
\textrm{s. t.}&&\frac{f_1(\mathrm{\boldsymbol{\delta_s}},\Theta_2,\boldsymbol{\alpha}_t,\boldsymbol{\alpha}_k)}{g_1(\mathrm{\boldsymbol{\delta_s}},\Theta_2,\boldsymbol{\alpha}_t,\boldsymbol{\alpha}_k)}\leq 1,\quad  \frac{f_2(\mathrm{\boldsymbol{\delta_s}},\Theta_2,\boldsymbol{\alpha}_t,\boldsymbol{\alpha}_k)}{g_2(\mathrm{\boldsymbol{\delta_s}},\Theta_2,\boldsymbol{\alpha}_t,\boldsymbol{\alpha}_k)}\leq 1, \quad \frac{f_3(\mathrm{\boldsymbol{\delta_s}},\Theta_2)}{g_3(\mathrm{\boldsymbol{\delta_s}},\Theta_2)}\leq 1 \\
&& -\sqrt{\frac{P_i}{P}}+ c_i \leq \delta_i \leq \sqrt{\frac{P_i}{P}}+ c_i \:\:,\: i =a, b\\
&& \boldsymbol{\delta}_s \in \mathbb{R}^2,\:\: \dv c \in \mathbb{R}^2, \:\: \Theta_2 \in \mathbb{R}.
\end{eqnarray}
\label{equivalent-form-optimizing-betas-for-given-integer-coefficients-B2}
\end{subequations}

The constraints \eqref{constraints-equivalent-form-optimizing-betas-for-given-integer-coefficients-B2} correspond to the constraints \eqref{PACF1}, \eqref{PACF2}, and \eqref{PACF3}. These functions consist of ratios of posynomials, i.e., are not posynomial --- recall that a ratio of posynomials is in general non posynomial. As before, we transform the complementary GP problem into a GP problem using series of approximations.

As there are several variables to optimize over simultaneously, the optimization is carried out in two steps. First, we optimize the power value $\mathrm{\boldsymbol{\delta_s}}$ using geometric programming with successive convex approximation in a way that is essentially similar to the above, for a fixed value of $\boldsymbol{\alpha}_t$ and $\boldsymbol{\alpha}_k$. Next, we optimize the value of $\boldsymbol{\alpha}_t$ and $\boldsymbol{\alpha}_k$ for a fixed value of $\mathrm{\boldsymbol{\delta}_s}$. This process is repeated until convergence. More specifically, the problem of finding the appropriate preprocessing vector $\boldsymbol{\delta}_s$ for a given integer vectors $\dv k$ and $\dv t$ can be solved using Algorithm B-2. 

\begin{algorithm}[t!]
\renewcommand{\thealgorithm}{B-2}
{\fontsize{9}{9}\selectfont
\caption{\small{Power allocation policy for $R^{\text{CoD}}_{\text{sym}}$ as given by (\ref{optimization-problem-compress-and-forward-at-relay-and-compute-at-destination})}}
\begin{algorithmic}[1]
\State Set $\mathrm{\boldsymbol{\delta_s}}^{(0,0)}$ to some initial value and set $\iota_2 =1$ and $\iota_3 =0$
\State Compute $\Theta_2^{(\iota_2-1,\iota_3)}$, $\boldsymbol{\alpha}_t^{(\iota_2-1,\iota_3)}$ and $\boldsymbol{\alpha}_k^{(\iota_2-1,\iota_3)}$ using $\boldsymbol{\delta_s}^{(\iota_2-1,\iota_3)}$.
\State Approximate $g(\boldsymbol{\delta_s}^{(\iota_2,\iota_3)},\Theta_2^{(\iota_2,\iota_3)})$ with $\tilde{g}(\boldsymbol{\delta_s}^{(\iota_2,\iota_3)},\Theta_2^{(\iota_2,\iota_3)})$ around $\boldsymbol{\delta_s}^{(\iota_2-1,\iota_3)}$ and $\Theta_2^{(\iota_2-1,\iota_3)}$ using \eqref{convex-approximations}
\State Solve the resulting approximated GP problem using an interior point approach. Denote the found solutions as $\mathrm{\boldsymbol{\delta_s}}^{(\iota_2,\iota_3)}$ and $\Theta_2^{(\iota_2,\iota_3)}$
\State Increment the iteration index as $\iota_2=\iota_2+1$ and go back to Step 3 using $\mathrm{\boldsymbol{\delta_s}}$ and $\Theta_2$ of step 4.
\State Terminate if $\|\mathrm{\boldsymbol{\delta_s}}^{(\iota_2,\iota_3)}-\mathrm{\boldsymbol{\delta_s}}^{(\iota_2-1,\iota_3)}\|\leq \epsilon_1$  and denote by $\boldsymbol{\delta}$ the final value
\State Increment the iteration index as $\iota_3=\iota_3+1$, set $\iota_2 =1$, and $\boldsymbol{\delta_s}^{(\iota_2-1,\iota_3)} = \boldsymbol{\delta}$ and then go back to Step 2
\State Terminate if $|R^{\text{CoD}}_{\text{sym}}[\iota_3]-R^{\text{CoD}}_{\text{sym}}[\iota_3-1]| \leq \epsilon_2$ 
\end{algorithmic}
}
\end{algorithm}

\section{Numerical Examples}\label{secV}

In this section, we provide some numerical examples. We measure the performance using symmetric-rate. We compare our coding strategies with those described in Section \ref{secII_subsecC}. 

Throughout this section, we assume that the channel coefficients are modeled with independent and randomly generated variables, each generated according to a zero-mean Gaussian distribution whose variance is chosen according to the strength of the corresponding link. More specifically, the channel coefficient associated with the link from Transmitter $A$ to the relay is modeled with a zero-mean Gaussian distribution with variance $\sigma^2_{\text{ar}}$; that from Transmitter $B$ to the relay is modeled with a zero-mean Gaussian distribution with variance $\sigma^2_{\text{br}}$; and that from the relay to the destination is modeled with a zero-mean Gaussian distribution with variance $\sigma^2_{\text{rd}}$. Similar assumptions and notations are used for the direct links from the transmitters to the destination. Furthermore, we assume that, at every time instant, all the nodes know, or can estimate with high accuracy, the values taken by the channel coefficients at that time, i.e., full channel state information (CSI). Also, we set $P_a = 20$ dBW, $P_b = 20$ dBW, $P_r = 20$ dBW and $P = 20$ dBW.

\begin{figure}[!ht]
  \begin{center}
  \includegraphics[scale=0.75]{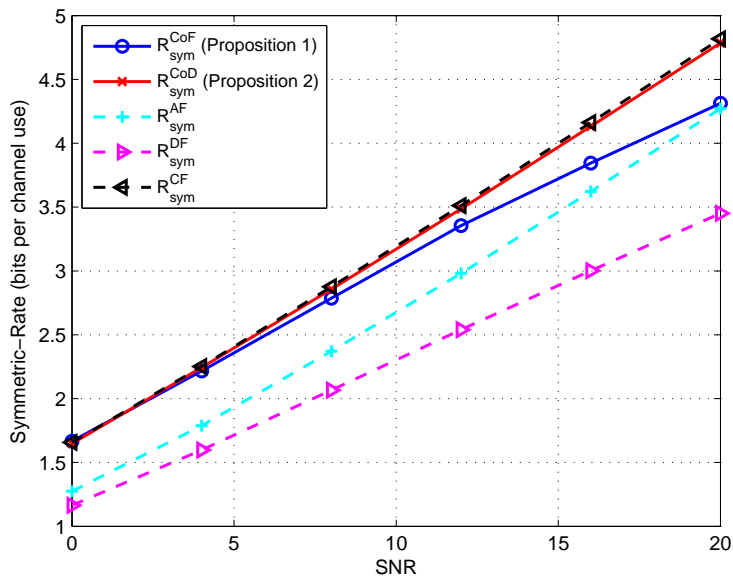}
  \end{center}
  \caption{Achievable symmetric rates. Numerical values are $P=20$ dBW, $\sigma^2_{\text{ar}}= 26$ dBW, $\sigma^2_{\text{br}}=26$ dBW, $\sigma^2_{\text{rd}}=18$ dBW, $\sigma^2_{\text{ad}}= 14$ dBW and $\sigma^2_{\text{bd}}=0$ dBW.}
\label{Fig2}
\end{figure}

Figure~\ref{Fig2} depicts the evolution of the symmetric-rate obtained using the so-called \textit{compute-and-forward at the relay} approach, i.e., the symmetric-rate $R^{\text{CoF}}_{\text{sym}}$ of proposition~\ref{proposition-achievable-sum-rate-compute-and-forward-at-relay} as given by \eqref{achievable-sum-rate-compute-and-forward-at-relay}; and the symmetric-rate obtained using the so-called \textit{compress-and-forward at the relay and compute at the destination} approach, i.e., the symmetric rate $R^{\text{CoD}}_{\text{sym}}$ of proposition~\ref{proposition-achievable-sum-rate-compress-and-forward-at-relay-and-compute-at-destination} as given by \eqref{achievable-sum-rate-compress-and-forward-at-relay-and-compute-at-destination}, as functions of the signal-to-noise ratio $\text{SNR}\:=10\log(P/N)$ (in decibels). Note that the curves correspond to numerical values of channel coefficients chosen such that $\sigma^2_{\text{ar}}= 26$ dBW, $\sigma^2_{\text{br}}=26$ dBW, $\sigma^2_{\text{rd}}=18$ dBW, $\sigma^2_{\text{ad}}= 14$ dBW and $\sigma^2_{\text{bd}}=0$ dBW. For comparison reasons, the figure also shows the symmetric rates obtained using the classical strategies of Section~\ref{secII_subsecC}, i.e., the symmetric-rate $R^{\text{AF}}_{\text{sym}}$ allowed by standard amplify-and-forward as given by \eqref{achievable-sum-rate-AF-symmetric-rate}, the symmetric-rate $R^{\text{DF}}_{\text{sym}}$ allowed by standard decode-and-forward as given by \eqref{achievable-sum-rate-DF-symmetric-rate}, and the symmetric-rate $R^{\text{CF}}_{\text{sym}}$ allowed by standard compress-and-forward as given by \eqref{achievable-sum-rate-CF-symmetric-rate}.

\begin{figure}[!ht]
  \begin{center}
  \includegraphics[scale=0.75]{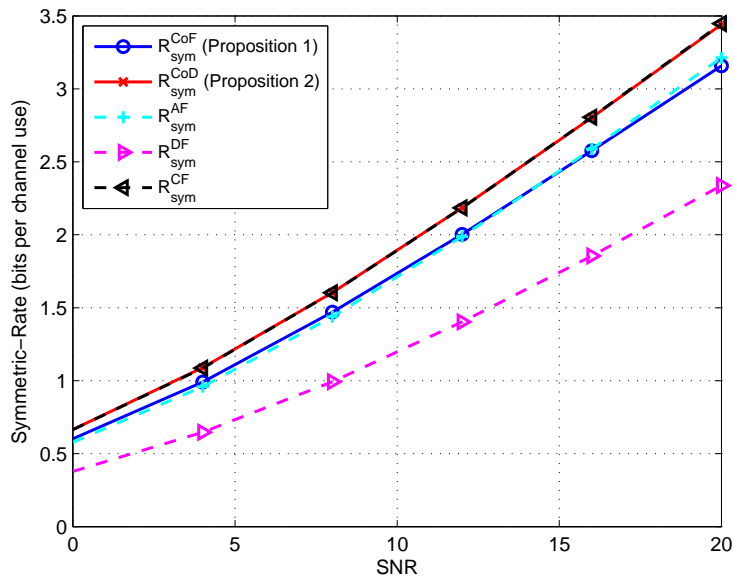}
  \end{center}
  \caption{Achievable symmetric rates. Numerical values are $P=20$ dBW, $\sigma^2_{\text{ar}}= 26$ dBW, $\sigma^2_{\text{br}}=0$ dBW, $\sigma^2_{\text{rd}}=26$ dBW, $\sigma^2_{\text{ad}}= 26$ dBW and $\sigma^2_{\text{bd}}=0$ dBW.}
\label{Fig3}
\end{figure}

For the example shown in Figure~\ref{Fig2}, we observe that the strategy of proposition~\ref{proposition-achievable-sum-rate-compress-and-forward-at-relay-and-compute-at-destination} achieves a symmetric-rate that is larger than what is obtained using standard DF and AF, and slightly less than what is obtained using standard CF (related to this aspect, recall the discussion in Remark~\ref{remark-local-computation-vs-CF}). Also, we observe that the strategy of proposition~\ref{proposition-achievable-sum-rate-compute-and-forward-at-relay} achieves a symmetric-rate that is larger than what is obtained using standard DF and AF. However, we observe that the strategy of proposition~\ref{proposition-achievable-sum-rate-compute-and-forward-at-relay} suffers from a loss in the degrees of freedom and that standard AF provides symmetric-rate that is better than the one obtained using the strategy of proposition~\ref{proposition-achievable-sum-rate-compute-and-forward-at-relay} at large $\text{SNR}$. This issue may be solved by incorporating binning via Slepian-Wolf at the relay (related to this aspect, recall the discussion in Remark~\ref{binning}).

Figure~\ref{Fig3} depicts the same curves for other combinations of channel coefficients, chosen such that $\sigma^2_{\text{ar}}= 26$ dBW, $\sigma^2_{\text{br}}=0$ dBW, $\sigma^2_{\text{rd}}=26$ dBW, $\sigma^2_{\text{ad}}= 26$ dBW and $\sigma^2_{\text{bd}}=0$ dBW. In this case, we observe that the strategy of proposition~\ref{proposition-achievable-sum-rate-compress-and-forward-at-relay-and-compute-at-destination} achieves a symmetric-rate that is as good as what is obtained using standard CF. Also, note that, the strategy of proposition~\ref{proposition-achievable-sum-rate-compute-and-forward-at-relay} provides a symmetric-rate that is slightly less than what is obtained using standard AF and is larger than what is obtained using standard DF.

\begin{remark}
Recall that, as we mentioned previously, the optimization ``Algorithm B" associated with the strategy of proposition~\ref{proposition-achievable-sum-rate-compress-and-forward-at-relay-and-compute-at-destination} is non-convex. In the figures shown in this paper, the symmetric rate provided by this strategy, which is based on Wyner-Ziv compression at the relay and computing appropriate equations at the destination, are obtained by selecting \textit{only} certain initial points for ``Algorithm B". For this reason, the symmetric-rate offered by the coding strategy of proposition~\ref{proposition-achievable-sum-rate-compress-and-forward-at-relay-and-compute-at-destination}, i.e., $R^{\text{CoD}}_{\text{sym}}$, can possibly be as good as the symmetric-rate offered by CF if one considers more initial points. Also, we note that even for those initial points choices which yield a symmetric rate that is slightly smaller than that obtained with standard compress-and-forward, there are advantages for using the coding scheme of proposition~\ref{proposition-achievable-sum-rate-compress-and-forward-at-relay-and-compute-at-destination} instead of regular CF, especially from a practical viewpoint as we already mentioned in Remark~\ref{remark-local-computation-vs-CF}. 
\end{remark}

\begin{figure}[!ht]
  \begin{center}
  \includegraphics[scale=0.75]{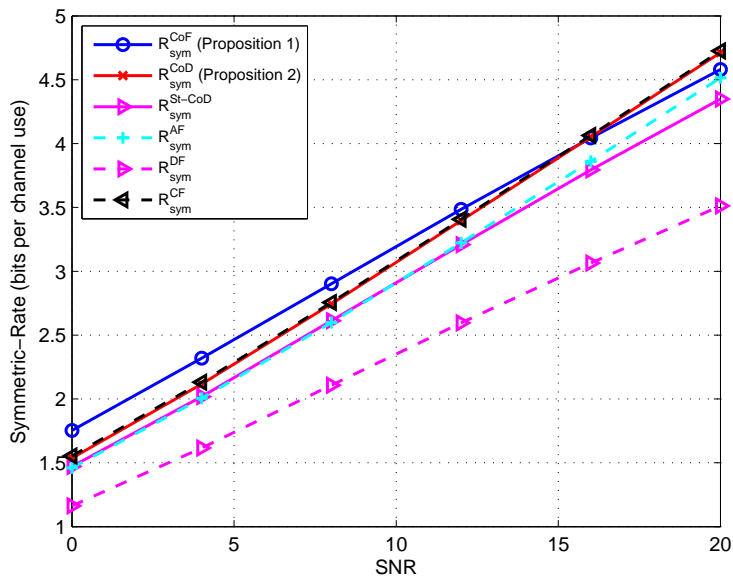}
  \end{center}
  \caption{Achievable symmetric rates. Numerical values are $P=20$ dBW, $\sigma^2_{\text{ar}}= 30$ dBW, $\sigma^2_{\text{br}}=18$ dBW, $\sigma^2_{\text{rd}}=15$ dBW, $\sigma^2_{\text{ad}}= 26$ dBW and $\sigma^2_{\text{bd}}=0$ dBW.}
  \label{Fig4}
\end{figure}

\begin{figure}[!ht]
  \begin{center}
  \includegraphics[scale=0.75]{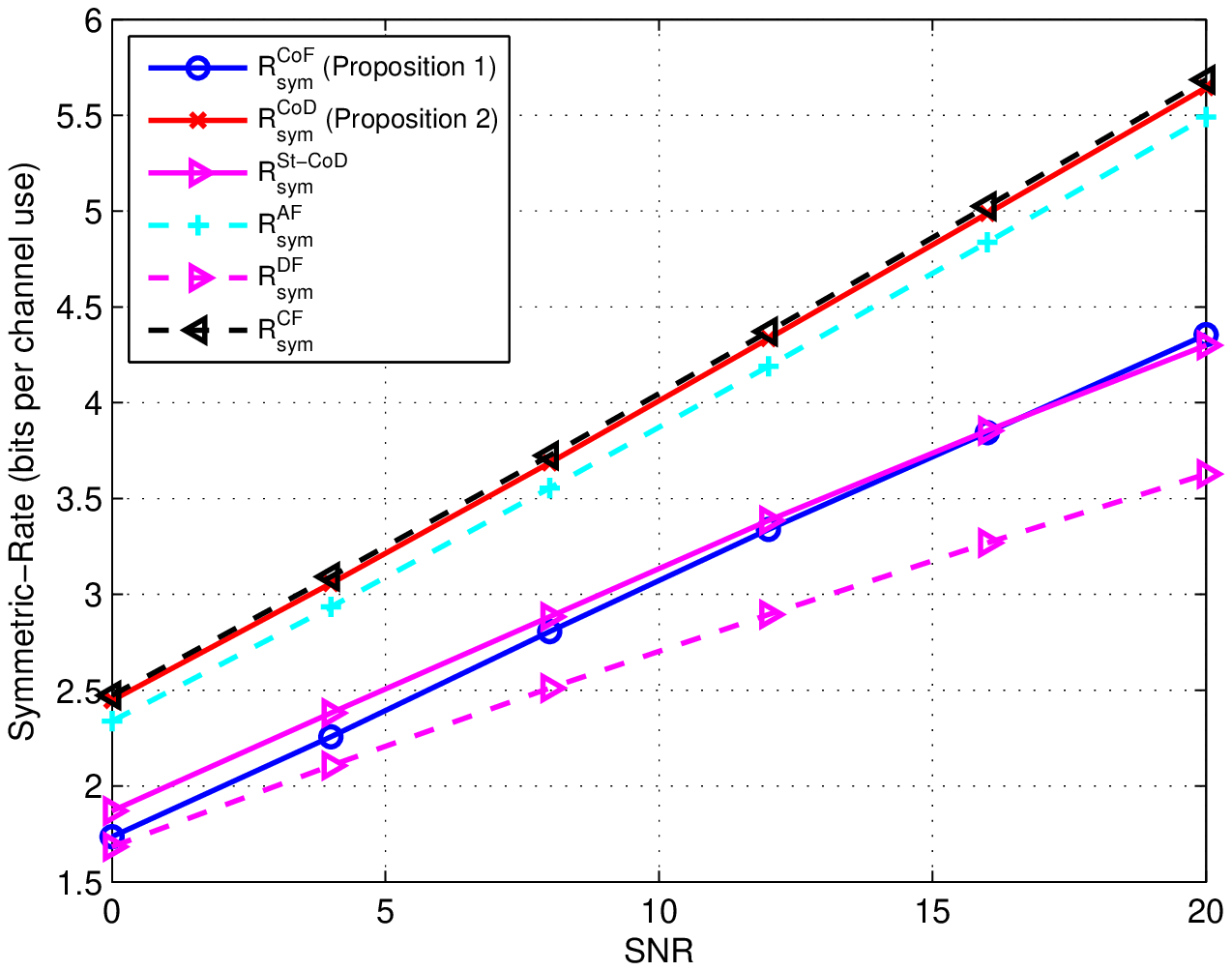}
  \end{center}
  \caption{Achievable symmetric rates. Numerical values are $P=20$ dBW, $\sigma^2_{\text{ar}}= 26$ dBW, $\sigma^2_{\text{br}}=26$ dBW, $\sigma^2_{\text{rd}}=14$ dBW, $\sigma^2_{\text{ad}}= 26$ dBW and $\sigma^2_{\text{bd}}=26$ dBW.}
  \label{Fig5}
\end{figure}

\begin{figure}[!ht]
  \begin{center}
  \includegraphics[scale=0.75]{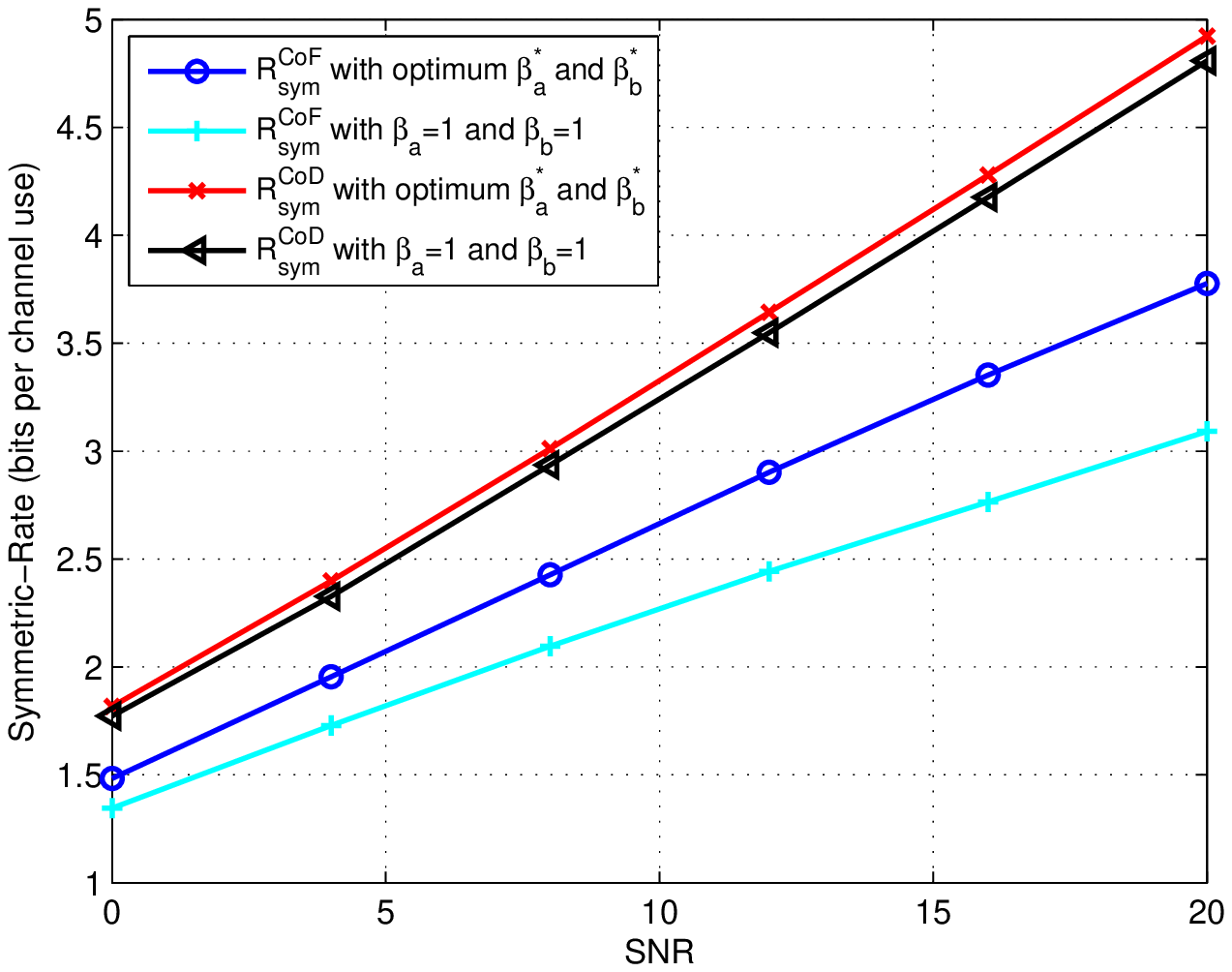}
  \end{center}
  \caption{Achievable symmetric rates. Numerical values are $P=20$ dBW, $\sigma^2_{\text{ar}}= 20$ dBW, $\sigma^2_{\text{br}}=20$ dBW, $\sigma^2_{\text{rd}}=20$ dBW, $\sigma^2_{\text{ad}}= 14$ dBW and $\sigma^2_{\text{bd}}=14$ dBW.}
  \label{Fig6}
\end{figure}

\begin{figure}[!ht]
  \begin{center}
  \includegraphics[scale=0.75]{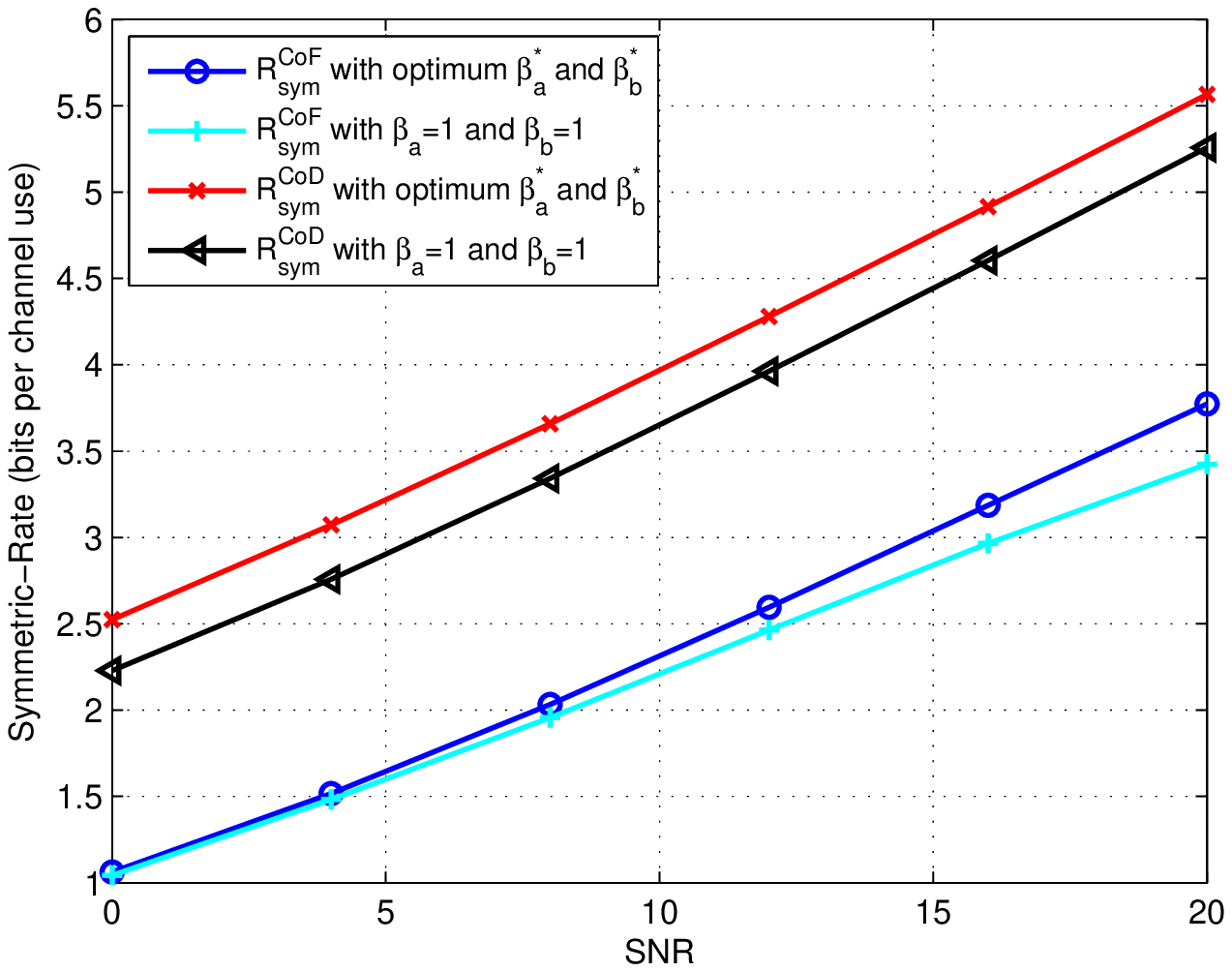}
  \end{center}
  \caption{Achievable symmetric rates. Numerical values are $P=20$ dBW, $\sigma^2_{\text{ar}}= 26$ dBW, $\sigma^2_{\text{br}}=26$ dBW, $\sigma^2_{\text{rd}}=0$ dBW, $\sigma^2_{\text{ad}}= 26$ dBW and $\sigma^2_{\text{bd}}=26$ dBW.}
  \label{Fig7}
\end{figure}

\begin{remark}
The comparison of the coding strategies of proposition~\ref{proposition-achievable-sum-rate-compute-and-forward-at-relay} and proposition~\ref{proposition-achievable-sum-rate-compress-and-forward-at-relay-and-compute-at-destination} is insightful. Generally, none of the two coding schemes outperforms the other for all ranges of $\text{SNR}$, and which of the two coding schemes performs better depends on both the operating $\text{SNR}$ and the relative strength of the links. For example, observe that while the strategy of proposition~\ref{proposition-achievable-sum-rate-compress-and-forward-at-relay-and-compute-at-destination} outperforms that of proposition~\ref{proposition-achievable-sum-rate-compute-and-forward-at-relay} in the examples shown in Figures~\ref{Fig2} and \ref{Fig3}, the situation is reversed for the example shown in Figure~\ref{Fig4} for some SNR ranges (related to this aspect, recall the discussion in Remark~\ref{remark-local-vs-distributed-computation}). For comparison reasons, the figure also shows the symmetric rate, $R^{\text{St-CoD}}_{\text{sym}}$, obtained by modifying the coding strategy of proposition~\ref{proposition-achievable-sum-rate-compress-and-forward-at-relay-and-compute-at-destination}. In this scheme, i.e., the modified strategy of proposition~\ref{proposition-achievable-sum-rate-compress-and-forward-at-relay-and-compute-at-destination}, accounting for the side information available at the destination through the direct links, the relay compresses what it gets using Wyner-Ziv compression and conveys it to the destination. The destination recovers the compressed version of the relay's output sent by the relay during the second transmission period by utilizing its output as well as the available side information received during the first transmission period. However, by opposition to the strategy of proposition~\ref{proposition-achievable-sum-rate-compress-and-forward-at-relay-and-compute-at-destination}, the destination does not combine the output from the users' transmission during the first transmission period and the recovered compressed version of the relay's output. That is, the destination computes the first equation using \textit{only} the recovered compressed version of the relay's output, and the second equation using \textit{only} the direct transmissions from the transmitters during the first transmission period. In Figure~\ref{Fig4}, we observe that the symmetric-rate $R^{\text{CoD}}_{\text{sym}}$ always outperforms the symmetric-rate $R^{\text{St-CoD}}_{\text{sym}}$ and this is precisely due to the joint processing implemented at the destination. We also observe that the symmetric-rate $R^{\text{CoF}}_{\text{sym}}$ is larger than the symmetric-rate $R^{\text{St-CoD}}_{\text{sym}}$. However, the situation is reversed for the example shown in Figure~\ref{Fig5}.  
\end{remark}

Figure~\ref{Fig6} shows the symmetric-rate $R^{\text{CoF}}_{\text{sym}}$ of proposition~\ref{proposition-achievable-sum-rate-compute-and-forward-at-relay} with optimum preprocessing allocation $\boldsymbol{\beta}^*$; the symmetric-rate $R^{\text{CoD}}_{\text{sym}}$ of proposition~\ref{proposition-achievable-sum-rate-compress-and-forward-at-relay-and-compute-at-destination} with optimum preprocessing allocation $\boldsymbol{\beta}^*$; the symmetric-rate $R^{\text{CoF}}_{\text{sym}}$ of proposition~\ref{proposition-achievable-sum-rate-compute-and-forward-at-relay} with no preprocessing allocation, i.e., $\boldsymbol{\beta} = \dv 1$; and the symmetric-rate $R^{\text{CoD}}_{\text{sym}}$ of proposition~\ref{proposition-achievable-sum-rate-compress-and-forward-at-relay-and-compute-at-destination} with no preprocessing allocation, i.e., $\boldsymbol{\beta} = \dv 1$. We observe that the strategy of proposition~\ref{proposition-achievable-sum-rate-compute-and-forward-at-relay} with optimum preprocessing vector $\boldsymbol{\beta}^*$ offers significant improvement over the one with no preprocessing allocation, and this improvement increase with the SNR. We also observe that the strategy of proposition~\ref{proposition-achievable-sum-rate-compress-and-forward-at-relay-and-compute-at-destination} with optimum preprocessing vector $\boldsymbol{\beta}^*$ offers small improvement over the one with no preprocessing allocation. However, with different numerical values of channel coefficients, we observe in Figure~\ref{Fig7} that the strategy of proposition~\ref{proposition-achievable-sum-rate-compress-and-forward-at-relay-and-compute-at-destination} with optimum preprocessing vector $\boldsymbol{\beta}^*$ offers significant improvement over the one with no preprocessing allocation.

\begin{figure}[!ht]
  \begin{center}
  \includegraphics[scale=0.75]{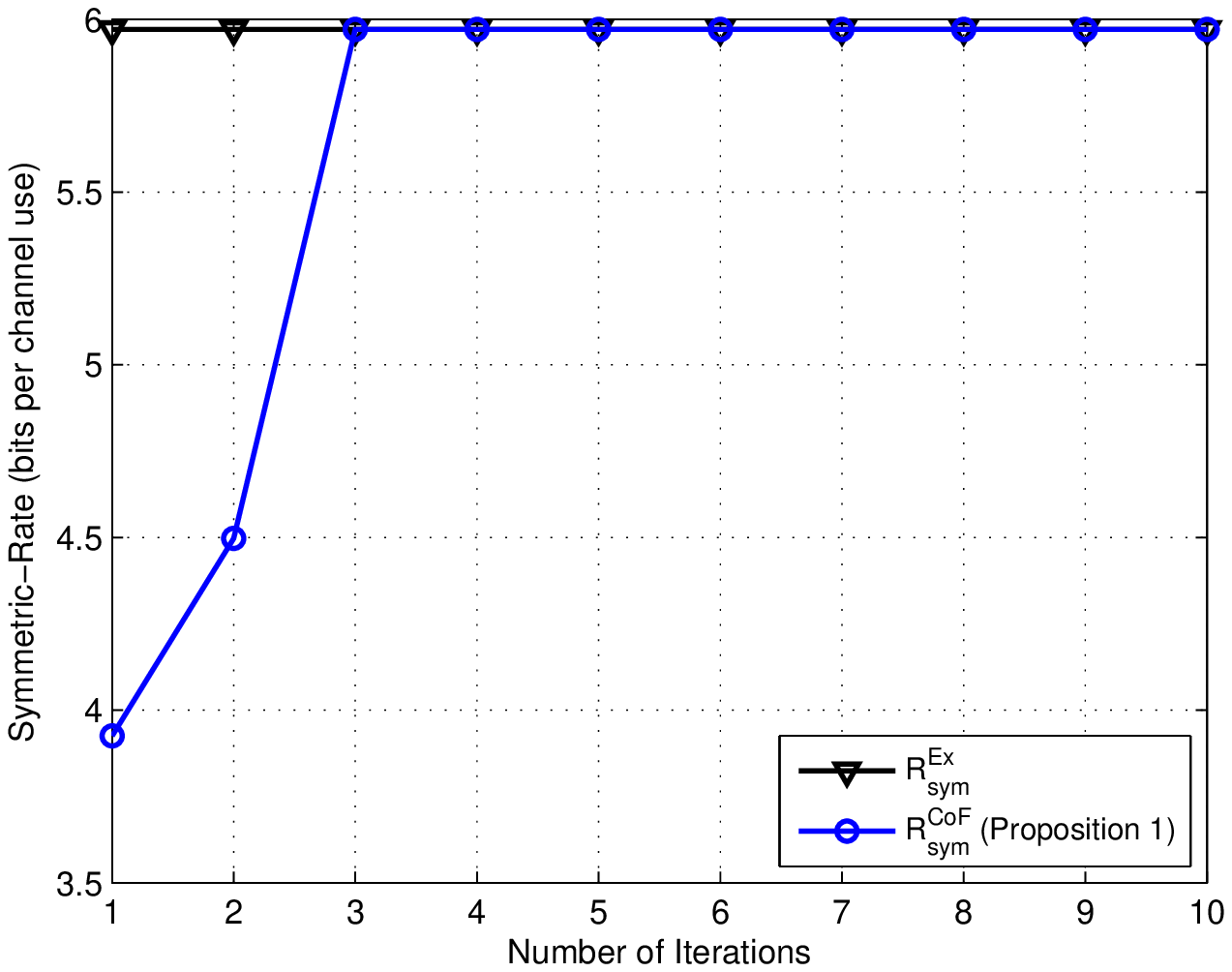}
  \end{center}
  \caption{Algorithm A v.s. Exhaustive search. Numerical values are $P=20$ dBW, SNR$=15$dB, $\sigma^2_{\text{ar}}=\sigma^2_{\text{br}}=\sigma^2_{\text{rd}}=20$ dBW, and $\sigma^2_{\text{ad}}=\sigma^2_{\text{bd}}=0$ dBW.}
\label{Fig8}
\end{figure}

We close this section with a brief discussion of the convergence speed of Algorithm A that we use to solve the optimization problem (A) given by \eqref{statement-optimization-problem-first-strategy}, as described in section \ref{secIV_subsecA}. Recall that the algorithm involves allocating the integer coefficients and the users' powers alternately, in an iterative manner. For a given set of powers, we find the best integer coefficients by solving a MIQP problem with quadratic constraints using the optimization software MOSEK. For a given set of integer-valued coefficients, we find the best powers at the sources and relay by solving a series of geometric programs by means of an interior point approach \cite{C05}. 

In order to investigate the convergence speed of the proposed algorithm, we compare it with one in which the integer coefficients search is performed in an exhaustive manner and the power allocation is kept as in Section~\ref{secIV_subsecA_subsubsec3}. Note that, using this exhaustive-search algorithm, for the integer valued equations coefficients to be chosen optimally, the search can be restricted to the set of integer values that satisfy $||\dv k||^2 \leq 1+||\dv h_r||^2\text{snr}$ and $||\dv t||^2 \leq 1+||\dv h_d||^2\text{snr}$, since otherwise the allowed symmetric rate is zero \cite{ZNGE09}. Let $R^{\text{Ex}}_{\text{sym}}$ denote the symmetric rate obtained by using the described exhaustive search-based algorithm. Figure~\ref{Fig8} shows that the number of iterations required for Algorithm A to converge, i.e., yield the same symmetric-rate as the one obtained through exhaustive search, is no more than three. Also, we note that, in comparison, the exhaustive search-based algorithm is more largely time- and computationally resources consuming, especially at large values of $\text{SNR}$. Similar observations, that we omit here for brevity, also hold for Algorithm B.

\section{Conclusion}\label{secVI}

In this paper, we study a two-user half-duplex multiaccess relay channel. Based on Nazer-Gastpar compute-and-forward scheme, we develop and evaluate the performance of coding strategies that are of network coding spirit. In this framework, the destination does not decode the information messages directly from its output, but uses the latter to first recover two linearly independent integer-valued combinations that relate the transmitted symbols. We establish two coding schemes. In the first coding scheme, the two required linear combinations are computed in a distributive manner: one equation is computed at the relay and then forwarded to the destination, and the other is computed directly at the destination using the direct transmissions from the users. In the second coding scheme, the two required linear combinations are both computed locally at the destination, in a joint manner. In this coding scheme, accounting for the side information available at the destination through the direct links, the relay compresses what it gets from the users using Wyner-Ziv compression and conveys it to the destination. The destination then computes the desired two linear combinations, locally, using the recovered output at the relay, and what it gets from the direct transmission from the users. For both coding schemes, we discuss the design criteria and establish the associated computation rates and the allowed symmetric rate. Next, for each of the two coding schemes, we investigate the problem of allocating the powers and the integer-valued coefficients of the recovered equations in a way to maximize the offered symmetric rate. This problem is NP hard; and in this paper we propose an iterative solution to solve this problem, through a careful formulation and analysis. For a given set of powers, we transform the problem of finding the best integer coefficients into a mixed-integer quadratic programming problem with quadratic constraints. Also, for a given set of integer-valued coefficients, we transform the problem of finding the best powers at the sources and the relay into series of geometric programs. Comparing our coding schemes with classic relaying techniques, we show that for certain channel conditions the first scheme outperforms standard relaying techniques; and the second scheme, while relying on feasible structured lattice codes, can offer rates that are as large as those offered by regular compress-and-forward for the multiaccess relay network that we study.

\section*{Acknowledgment}

The authors would like to thank Prof. Stephen Boyd and his team, from Information Systems Laboratory, Stanford University, USA, and MOSEK Aps for partly helping in the software and programs used in the optimization part in this paper. Also, the authors thank the anonymous reviewers for the relevant comments which helped improve the quality of this manuscript.



\bibliographystyle{unsrt}
\bibliography{bibfile}

\end{document}